\documentclass[11pt,a4paper]{article}
\pdfoutput=1 
\usepackage{jheppub2}
\usepackage{graphicx}  
\usepackage{dcolumn}   
\usepackage{amsmath}   
\usepackage{amssymb}   
\usepackage{cancel}
\usepackage{comment}
\usepackage[caption=false]{subfig}
\usepackage[utf8x]{inputenc}
\usepackage{newunicodechar}
\usepackage{color}
\usepackage{xcolor}
\usepackage{slashed}
\usepackage[caption=false]{subfig}
\definecolor{urlblue}{rgb}{0.2,0.4,0.7}
\definecolor{citegreen}{rgb}{0,0.6,0.2}
\definecolor{linkred}{rgb}{0.9,0.2,0.1}
\usepackage{hyperref}
\hypersetup{
colorlinks=true, citecolor=citegreen, linkcolor=blue, urlcolor=urlblue}
\usepackage{float}
\usepackage[mathscr]{euscript}
\usepackage{url}
\usepackage{cleveref}

\def\d{\ensuremath{\partial}}

 \definecolor{X575}{rgb}{0.05, 0.7, 0.05}

\begin{document}

\title{Probing non-standard $HVV (V=W, Z)$ couplings in single Higgs production at future electron-proton collider}
\author{Pramod Sharma,} 
\emailAdd{pramodsharma.iiser@gmail.com}
\affiliation{Indian Institute of Science Education and Research, Knowledge City, Sector 81, S. A. S. Nagar,Manauli PO 140306, Punjab, India.}\author{Ambresh Shivaji}
\emailAdd{ashivaji@iisermohali.ac.in}

\abstract{
The couplings of the Higgs boson ($H$) with massive gauge bosons of weak interaction ($V= W, Z$), can be probed in single Higgs boson production at 
the proposed future Large Hadron-Electron Collider (LHeC). In the collision of an electron with a proton, single Higgs production takes place via so-called charged-current ($e^-p \to \nu_e H j$) and neutral-current 
($e^-p \to e^-H j$) processes. We explore the potential 
 of the azimuthal angle correlation between the forward jet and scattered neutrino or electron  
in probing the  non-standard $HVV$ couplings at the collider center-of-mass energy of $\sqrt{s} \approx 1.3$~TeV. We choose the most general modifications (of $CP$-even and $CP$-odd nature) to these couplings due to new physics effects beyond the standard model.  
We derive exclusion limits on new physics parameters of $HVV$ couplings as a function of 
 integrated luminosity at $95$\% C.L. using the azimuthal angular correlations in charged- and neutral-current processes. We find that using 1000 $fb^{-1}$ data, the standard model-like new physics parameters in $HWW$ and $HZZ$ couplings can be constrained with accuracies of 4\% and 15\%, respectively. The least constrained $CP$-even parameters of $HWW$ coupling can be as large as 0.04, while those of $HZZ$ coupling can have values around 0.31.
Allowed values of $CP$-odd parameters in $HWW$ and $HZZ$ couplings are found to be around 0.14 and 0.34, respectively. We also 
 study changes in the allowed values of non-trivial new physics parameters in presence of other parameters.     
}

\maketitle

%
\section{Introduction}
\label{sec:intro}
The minimal electroweak standard model of particle physics predicts the existence of 
a fundamental massive scalar particle, the Higgs boson~\cite{GLASHOW:1961579,Englert:13321,HIGGS:1964132,Higgs:13508,Guralnik:13585,Weinberg:191264,Salam:xyz}. The mass of the Higgs boson 
is not a prediction of the model. However, once its mass is measured, its couplings 
with other standard model particles can be determined. The ATLAS and CMS experiments 
at the Large hadron Collider (LHC) have confirmed the discovery of a scalar particle of mass 
125 GeV which is very much like the standard model Higgs boson~\cite{ATLAS:2012yve,CMS:2012qbp,CMS:2013btf}. 
The fact that no clear evidence of new physics has emerged yet in the analyses 
of the LHC data on Higgs, precise measurement of various couplings in the Higgs sector is one 
of the main goals of the future high energy collider projects~\cite{deBlas:2019rxi}.
In a scenario
where high scale new physics effects in Higgs-vector boson couplings are parametrized by a common factor $\kappa_V$, 
the combined analyses of CMS and ATLAS taking the 
LHC Run-I data lead to $\kappa_V=1.03\pm0.03$~\cite{ATLAS:2015xyz,CMS:2015xyz}. The expected accuracy on $\kappa_V$ at the 
HL-LHC is below 2\%~\cite{Cepeda:2019klc}. It is well known that new physics effects may introduce new Lorentz structures and therefore new parameters in $HVV (V=W, Z)$ couplings~\cite{LHCHiggsCrossSectionWorkingGroup:2016ypw}.

The most general Lagrangian which can account for all possible 
three-point interactions involving Higgs and massive electroweak gauge bosons can be written as,
\begin{align}
	{\cal L}_{_{HVV}}^{\rm BSM}
	= &     g\ \left( m_V \kappa_W W^+_\mu W^{-\mu} + \dfrac{\kappa_Z}{2\,\cos\theta_W} m_Z Z_\mu Z^\mu \right) H
 \notag\\
	-  \frac{g}{m_W} \bigg[ & \dfrac{\lambda_{1W}}{2} W^{+\mu\nu} W^{-}_{\mu\nu} 
	+ \dfrac{\lambda_{1Z}}{4} Z^{\mu\nu} Z_{\mu\nu} 
 \notag\\
	+ & \lambda_{2W} ( W^{+\nu} \d^\mu W^{-}_{\mu\nu}  + {h.c.} ) +  \lambda_{2Z} Z^{\nu} \d^\mu Z_{\mu\nu}   
 \notag\\
	+ & \dfrac{\tilde{\lambda}_W}{2} W^{+\mu\nu} \widetilde{W}^{-}_{\mu\nu}   +  
	    \dfrac{\tilde{\lambda}_Z}{4} Z^{\mu\nu} \widetilde{Z}_{\mu\nu} \bigg] H,
 \label{lag1}
\end{align}
where $g$ is the $SU(2)$ coupling parameter and $\widetilde{V}^{\mu\nu} = \frac{1}{2} \epsilon^{\mu\nu\rho\sigma}V_{\rho\sigma}$ is the dual 
field strength tensor. The beyond the standard model (BSM) parameters $\kappa_V$ and $\lambda_{iV}(i=1,2)$ are associated with $CP$-even, while $\tilde{\lambda}_V$ 
are associated with $CP$-odd couplings of the Higgs with vector bosons. The standard model (SM) predictions correspond to $\kappa_V=1, 
\lambda_{iV}=0=\tilde{\lambda}_V$. The Lorentz structures of Eqs.~\eqref{lag1} can be derived from the $SU(2)_L \otimes U(1)_Y$ gauge 
invariant dimension-6 operators~\cite{BUCHMULLER1986621, Giudice:2007fh, Grzadkowski:2010es, Alloul:2013naa, Falkowski:2014tna, Brivio:2017vri}. The above framework is equivalent to the so-called Higgs basis~\cite{LHCHiggsCrossSectionWorkingGroup:2016ypw}. 
The $HVV$ vertex factor in our framework is given by,
\begin{align}
	\Gamma_{HVV}^{\mu\nu}(p_1,p_2) = & g_V m_V \kappa_V g^{\mu\nu} + \frac{g}{m_W} [\lambda_{1V} (p_1^\nu p_2^\mu - g^{\mu \nu} p_1.p_2) \notag \\
	& + \lambda_{2V} (p_1^{\mu}p_1^{\nu}+p_2^{\mu}p_2^{\nu}-g^{\mu \nu} p_1.p_1-g^{\mu \nu} p_2.p_2) \notag \\
	& +  \widetilde{\lambda}_V ~ \epsilon^{\mu \nu \alpha \beta} p_{1 \alpha} p_{2 \beta} ].
\end{align}
Here, $p_1$ ($\mu$) and $p_2$ ($\nu$) denote the momenta (Lorentz indices) of the two vector bosons in $HVV$ coupling. Also, we have defined
$g_W=g$ and $g_Z = g/{\rm cos}\theta_W$, $\theta_W$ being the Weinberg angle. Note that the parameter $\lambda_{2V}$ 
is linked with the off-shellness of the vector bosons.

The new physics parametrization similar to the above one has been used to study $HVV$ couplings at various current and future colliders~\cite{Hagiwara:1993sw, Hagiwara:2000tk, Han:2000mi, Han:2005pu, Biswal:2008tg, Dutta:2008bh, Christensen:2010pf, Desai:2011yj, Biswal:2012mp,  Cakir:2013bxa, Maltoni:2013sma, Anderson:2013afp, Kumar:2015kca, Boselli:2017pef, Nakamura:2017ihk, Li:2019evl, Sahin:2019wew, Banerjee:2019pks, Han:2020pif, Bizon:2021rww, Rao:2019hsp, Rao:2022olq, Asteriadis:2022ebf}. Several studies exist in the literature which consider new physics effects in $HVV$ couplings in an EFT framework via dimension-6 operators~\cite{Senol:2012fc, Banerjee:2013apa, Craig:2014una,Amar:2014fpa, Englert:2014cva, Ellis:2014dva, Mellado:2015ehl, Banerjee:2015bla, Dwivedi:2015nta, Englert:2015hrx, Craig:2015wwr, Dwivedi:2016xwm, Ferreira:2016jea, Degrande:2016dqg, Denizli:2017pyu,  Khanpour:2017cfq,Hesari:2018ssq, Banerjee:2018bio, Karadeniz:2019upm, Cirigliano:2019vfc, Freitas:2019hbk, Henning:2019vjr, Biswas:2021qaf}. For a one-to-one correspondence between the two frameworks dictionaries like~\cite{Boselli:2017pef} can be used.

In this paper, we study the effect of the BSM parameters of $HVV$ vertex in charged-current and neutral-current processes at the 
future Large Hadron Electron Collider (LHeC)~\cite{LHeCStudyGroup:2012zhm,Bruening:2013bga}. These processes have
been studied in the context of Higgs boson searches in $H \to b{\bar b}$ decay mode~\cite{Han:2009pe}. The role of azimuthal angle between missing energy and jet in the transverse plane was investigated to distinguish $CP$-odd coupling from the $CP$-even coupling in the charged-current process~\cite{Biswal:2012mp}. In the present study,
we have extended the theoretical framework considered in~\cite{Biswal:2012mp} which takes note of the fact that in charged-current and neutral-current processes the virtuality of mediating $W$ and $Z$ bosons changes event-by-event. This effect introduces one more $CP$-even parameter $\lambda_{2V}$ in the analysis.
We also study the neutral current process in detail using the azimuthal angle correlation between the final state $e^-$ and the jet.

\section{BSM effects in Single Higgs production}
\label{sec:bsm}

\begin{figure}
\begin{center}
\subfloat[]{\includegraphics[width=0.40\textwidth]{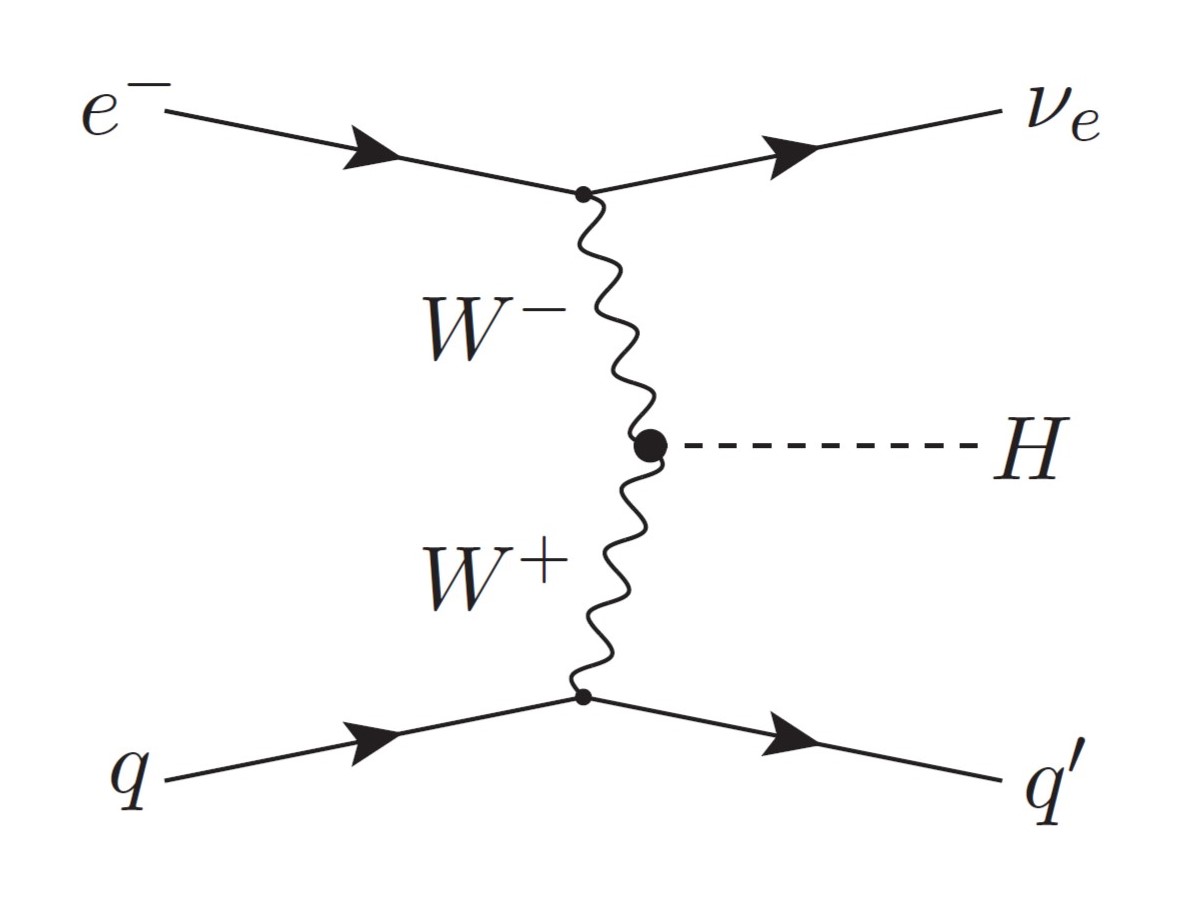}} \hspace{2cm}
\subfloat[]{\includegraphics[width=0.40\textwidth]{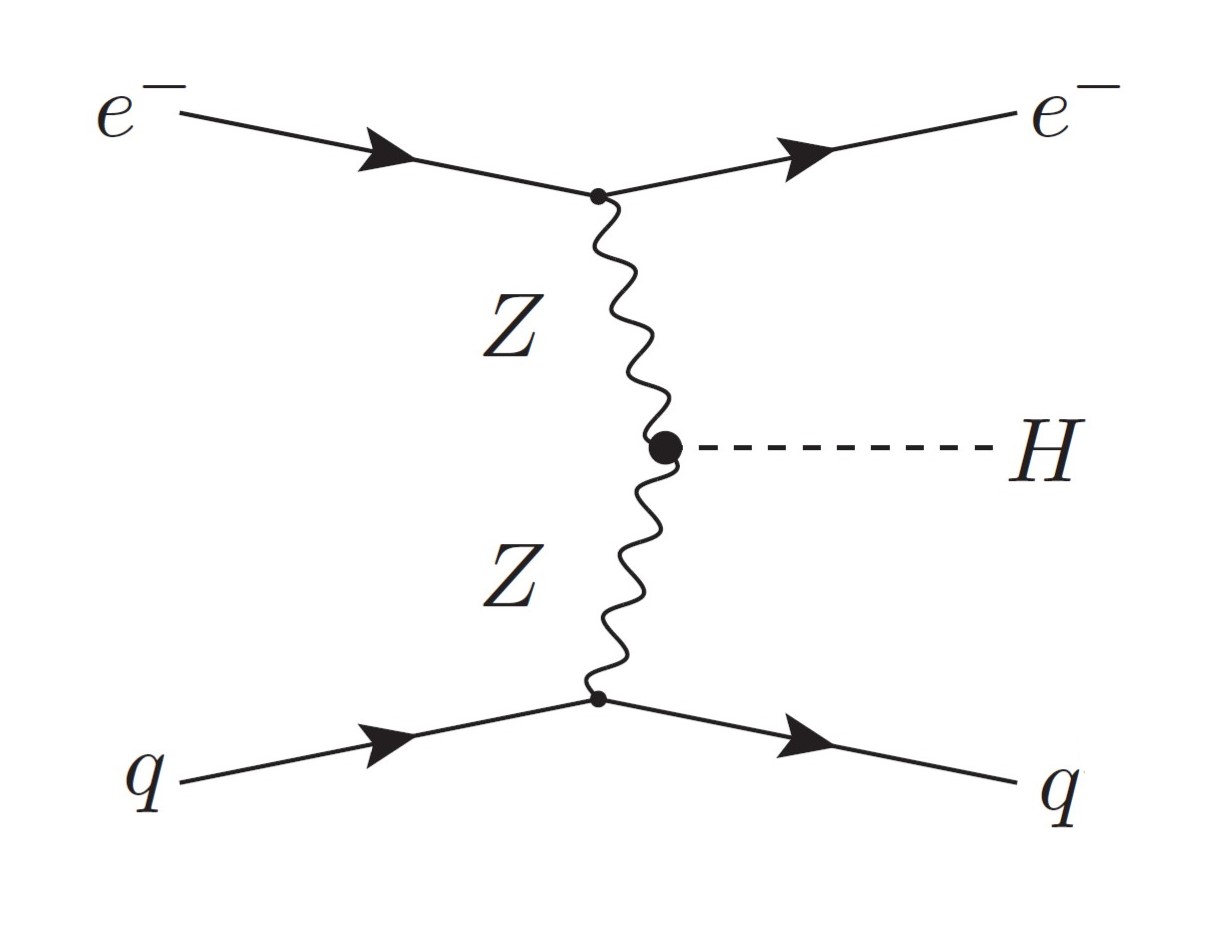}}
	\caption{Leading order Feynman diagrams for single Higgs production at $e^- p$ collider. $(a)$ Charged-current process. $(b)$ Neutral-current process.}
\label{fig:figW}
\end{center}
\end{figure}

At $e^- p$ collider, the single Higgs production takes place via charged-current (\texttt {CC}) and neutral-current (\texttt {NC}) processes,
\begin{eqnarray*}
e^-p &\to& \nu_e H j + X \hspace{1.45cm}[\texttt{CC}] \\
e^-p &\to& e^- H j + X. \hspace{1.30cm}[\texttt{NC}]
\end{eqnarray*}
Leading order Feynman diagrams for these processes are shown in figure~\ref{fig:figW}. The {\tt CC} and {\tt NC} processes are sensitive to 
$HWW$ and $HZZ$ couplings, respectively.
These processes are very similar to the vector boson fusion (VBF) 
processes $pp \to Hjj$ for Higgs production at the LHC. However, unlike at the LHC, there is no contamination from $pp \to HW, HZ \to Hjj$ type 
higstrahlung processes at $e^- p$ collider. Further, if we compare $e^- p \to e^-Hj$ with $pp \to Hjj$, we find that due to an asymmetry in the beam type 
and beam energies, the rapidity difference $\Delta \eta(e,j)$ distribution is shifted towards the left with respect to $\Delta \eta(j,j)$. Also, the Higgs produced in $e^- p$ collisions 
is always in the forward or backward region while the Higgs produced in $pp$ collisions, due to a symmetric environment, is detected in both forward 
and backward regions. 

\begin{figure}
\centering
\subfloat[]{\includegraphics[width=0.50\textwidth]{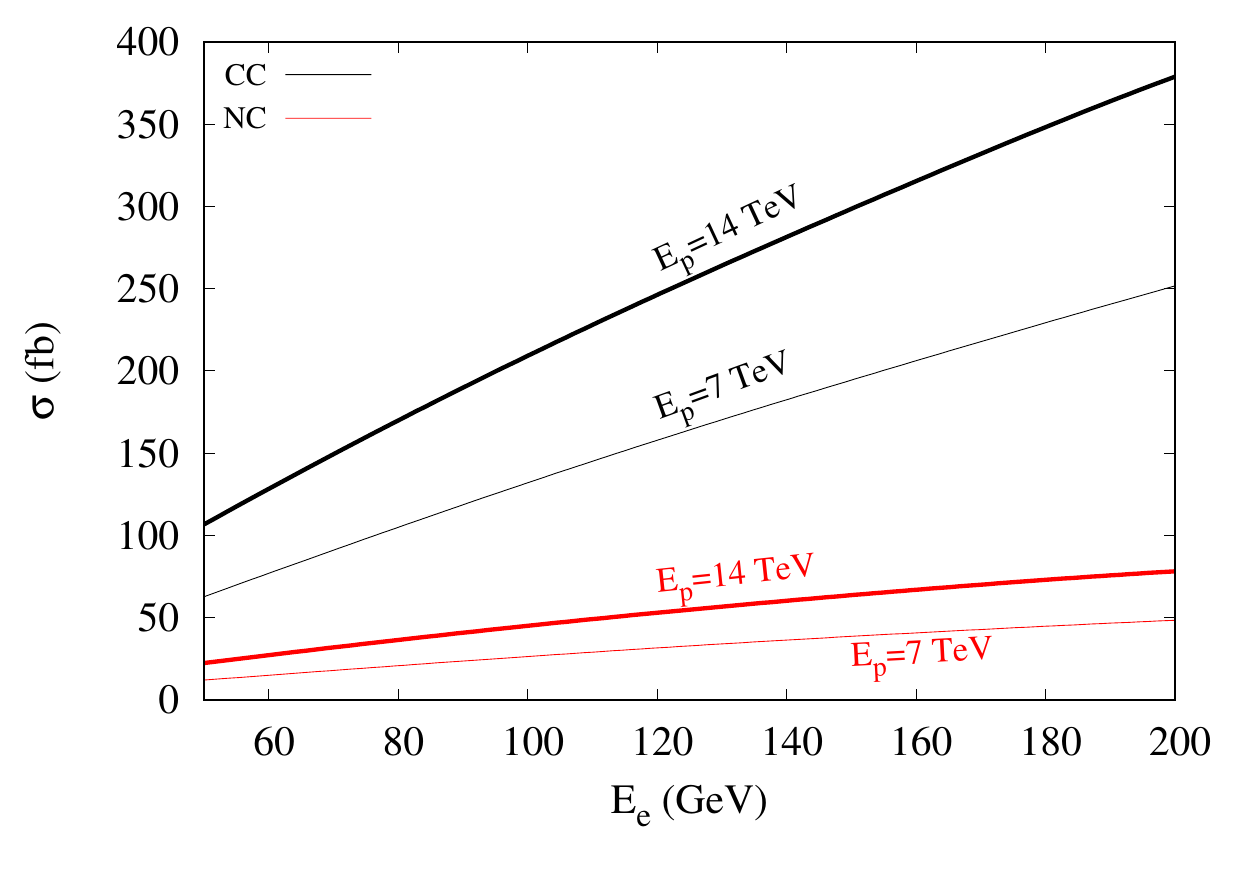}}
\subfloat[]{\includegraphics[width=0.50\textwidth]{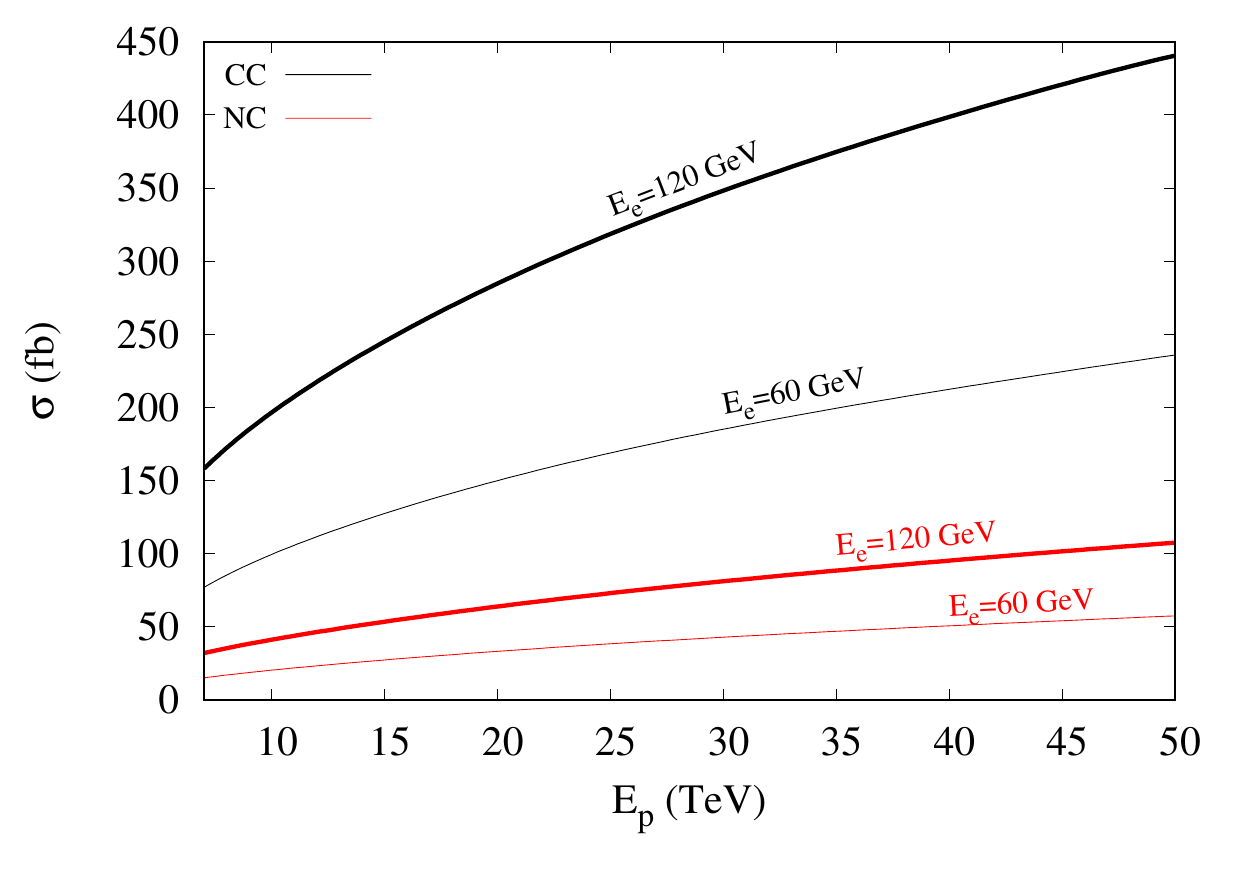}}
	\caption{Variation of {\tt CC} (black) and {\tt NC} (red) cross sections with respect to the electron beam energy, $E_e$ (left), and proton beam energy, $E_p$ (right).}
\label{fig:XS-sm}
\end{figure}

\begin{figure}
\centering
\subfloat[]{\includegraphics[width=0.5\textwidth]{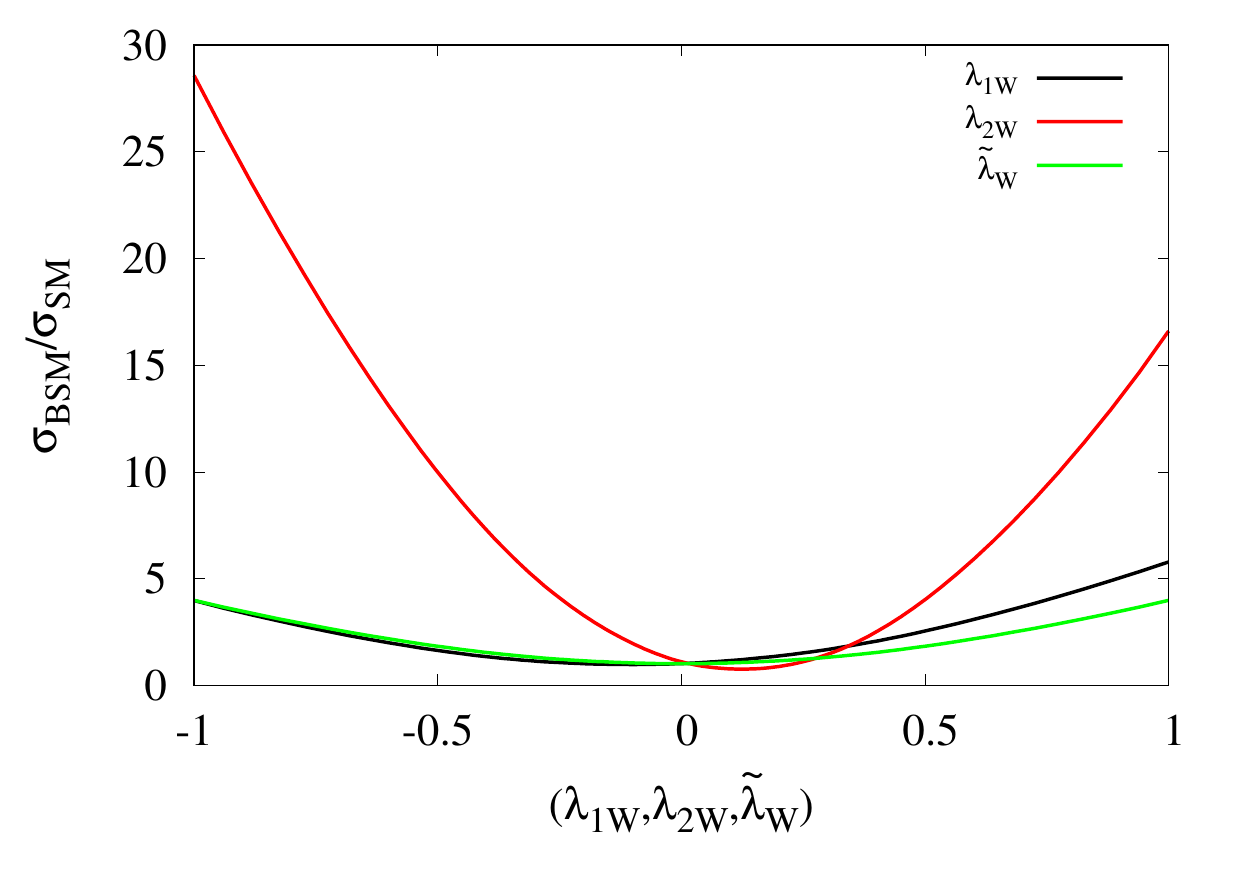}}
\subfloat[]{\includegraphics[width=0.5\textwidth]{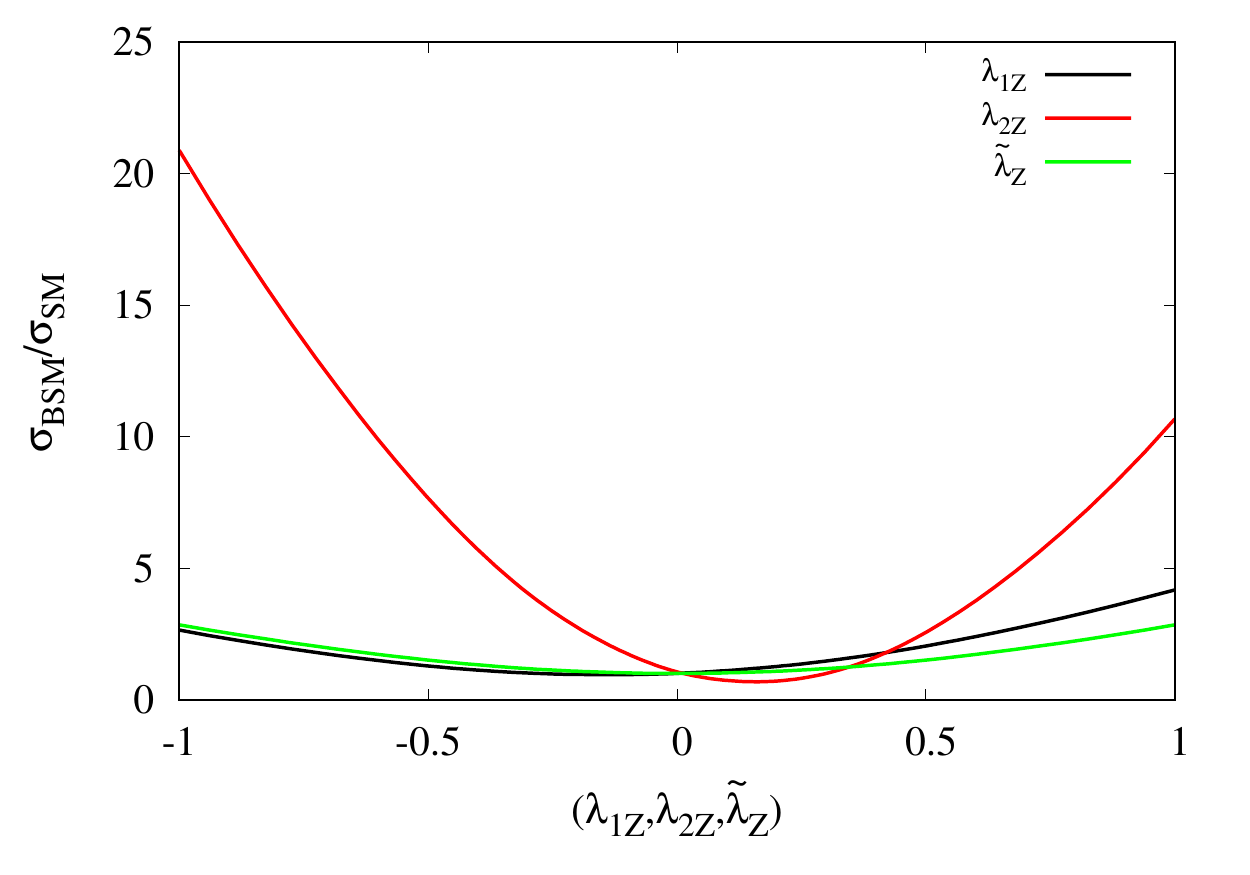}}
	\caption{Variation of {\tt CC} (left) and {\tt NC} (right)  cross sections with respect to the non-trivial BSM parameters.}
\label{fig:XS-bsm}
\end{figure}

In various proposals and studies for $e^- p$ collider, the electron beam energy has been considered in the 
range of 50-200 GeV~\cite{LHeC:2020van}. On 
the other hand, keeping the proposal of FCC-hh~\cite{Benedikt:2022kan} in mind, a proton beam with energy as large as 50 TeV can also be taken.  
In figure \ref{fig:XS-sm}, we show the dependence of the standard model cross section for possible electron and proton beam energies. 
The cross sections in these plots are produced with default cuts and settings in {\tt MadGraph5\_aMC@NLO}(MG5)~\cite{Alwall:2014hca}. The BSM model file for madgraph is produced using the {\tt FeynRules} package~\cite{Alloul:2013bka}.  
We note that the effect of electron beam energy on the cross section is stronger than that of proton beam energy. 
For our present study, we choose $E_e=60$ GeV and $E_p=7$ TeV which corresponds to a center-of-mass energy of 1.3 TeV. 
At this energy, the standard model cross sections for {\tt CC} and {\tt NC} processes 
are respectively 88 $fb$ and 16 $fb$ with the unpolarized electron beam. With the -80 \% polarized electron beam, the 
cross sections become 158 $fb$ and 19 $fb$ respectively. The significant change in the {\tt CC} cross section when using the 
polarized electron beam is simply related to the fact that the $W$ boson couples to the left-handed fermions only.

In presence of the BSM parameters introduced above, the inclusive or differential cross section, denoted by $X$, can be written symbolically as
\begin{eqnarray}
	X^{\rm BSM} = X^{\rm SM} + \sum_i X_i c_i + \sum_{i,j} X_{ij} c_i c_j.
	\label{xsection}
\end{eqnarray}
Here, $c_i= \kappa_V, \lambda_{1V}, \lambda_{2V}, {\tilde \lambda}_V$.  Variation of the cross sections with respect to each non-trivial BSM parameter is shown in 
figure~\ref{fig:XS-bsm} for both {\tt CC} and {\tt NC} processes. We can infer that the inclusive cross section is most sensitive to $\lambda_{2V}$ and least sensitive to $\tilde{\lambda}_V$. Since the vector bosons are connected to massless fermions, terms proportional to $p_i^{\mu}p_i^{\nu} (i=1,2)$ do not contribute. The parameter $\lambda_{2V}$, thus, directly probes the off-shellness of the vector bosons.
Also, the variations for +ve and -ve values suggest that the linear term is more relevant in presence of $\lambda_{2V}$ than in presence of 
$\lambda_{1V}$. Being $CP$ even observable, the cross section does not depend on $\tilde{\lambda}_V$ linearly. 
The effect of $\kappa_V$ is standard model-like and the cross sections scale as $\kappa_V^2$. 

\begin{figure}[htp]
\centering
	\subfloat[]{\includegraphics[width=0.5\textwidth,height=0.32\textwidth]{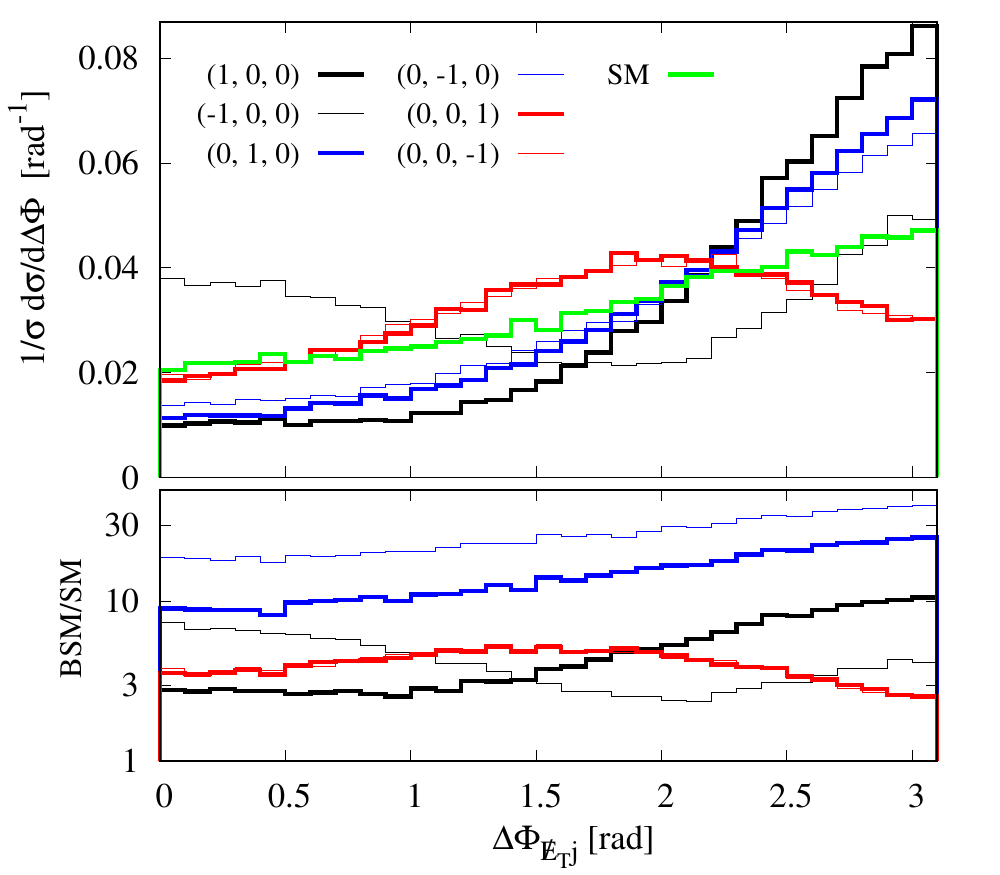}}
	\subfloat[]{\includegraphics[width=0.5\textwidth,height=0.32\textwidth]{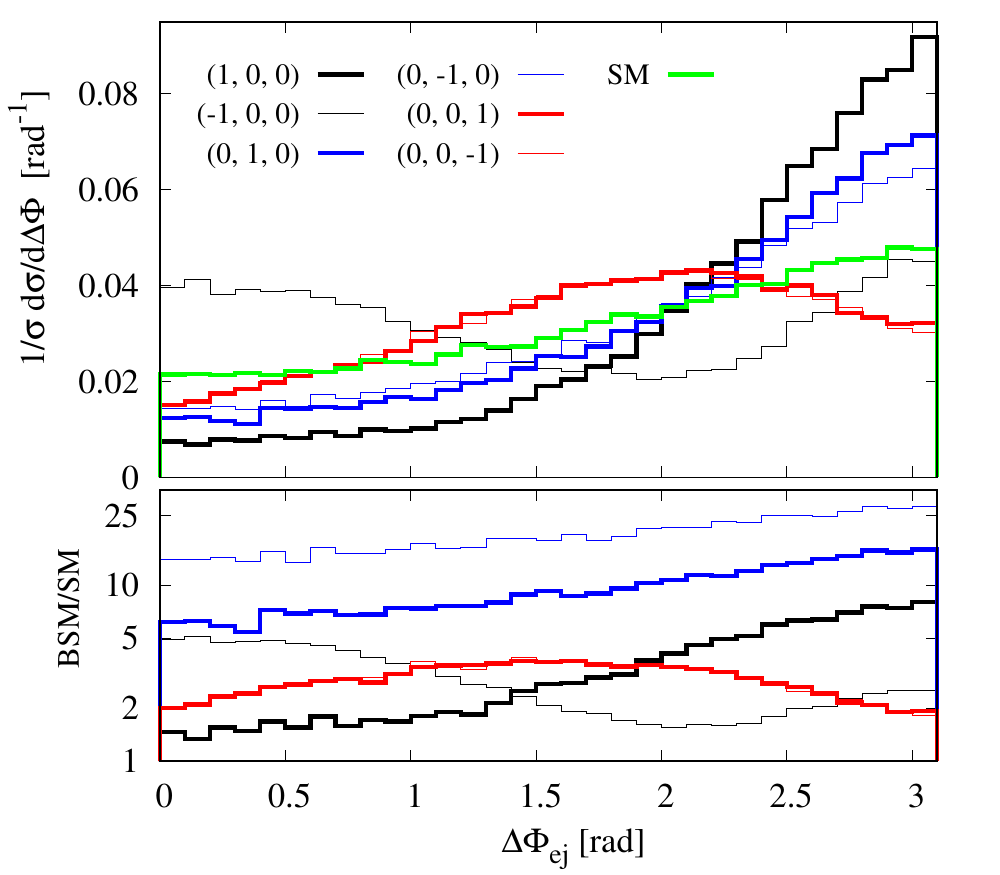}}
	\caption{ BSM effects in azimuthal angle distributions for {\tt CC} (left) and {\tt NC} (right) processes. The numerical values in the 
	brackets refer to the values of BSM parameters
	($\lambda_{1V}, \lambda_{2V}, {\tilde \lambda}_V $). The lower panel in each plot gives the ratio of BSM and SM predictions. }
	\label{fig:dphi}
\end{figure}

Our study involves both $CP$-even and $CP$-odd BSM parameters. It is, therefore, desirable to look for observables which can efficiently 
distinguish these two types of couplings. In the context of Higgs physics, pure $CP$-odd observables have been proposed in processes 
with charge-neutral final states at $e^+e^-$ and $pp$ colliders~\cite{Han:2000mi,Christensen:2010pf}. 
In Ref.~\cite{Biswal:2012mp} it has been shown that $\Delta \phi$ distribution (positive difference of azimuthal angles between $\slashed{E_T}$ 
and the jet) is useful in distinguishing $CP$-even coupling with $CP$-odd coupling 
 in the {\tt CC} process. A similar conclusion can be inferred from the VBF study at the LHC presented in~\cite{Maltoni:2013sma}.
 The behaviour of $\Delta \phi$ distributions (normalized by the total cross section) for individual parameters ($\lambda_{1V}$, $\lambda_{2V}$, and 
 $\tilde{\lambda}_V$) are shown in figure~\ref{fig:dphi} by taking 0, $\pm 1$ as benchmark values of these parameters. 
 The choices of benchmark values are completely ad hoc and these values are taken only to illustrate the individual effect of parameters 
 on $\Delta \phi$ distributions. As expected, the BSM effects are not flat across bins. We note that in both {\tt CC} and {\tt NC} processes, 
 $\Delta \phi$ for $CP$-odd parameter is characteristically different from $\Delta \phi$ for $CP$-even parameters. Also, the distribution 
 is symmetric for +ve and -ve values of $\tilde{\lambda}_V$. In presence of $\lambda_{1V}$, the distribution peaks in opposite direction 
 for +ve and -ve values. Thus $\Delta \phi$ distribution can differentiate between $\lambda_{1V}$ and $\lambda_{2V}$ as well. 

\section{Collider simulation: Signal vs Background}
\label{sec:S-vs-B}

Considering the dominant decay channel $ H \to b\bar{b}$, we identify 
signal events with  missing transverse energy (due to neutrinos), two $b$ jets, and 
one light jet $j=(u,d,c,s,g)$ in the final state of the {\tt CC} process. Apart from the QCD-induced irreducible backgrounds 
from $e^- p \to \nu_e b\bar{b} j$, there 
are a number of reducible backgrounds coming from; (i) $e^- p \to \nu_e \bar{t} b  \to \nu_e b \bar{b} j j $ in which the two jets are 
misidentified as one jet~\footnote{We have explicitly checked that other background channels with $\nu_e b \bar{b} j j$ final state 
are not significant after applying the analysis cuts discussed below.} (ii) $ e^- p \to  \nu_e j j j$ in which two of the three light 
jets are mistagged as $b$ jets.  
Another reducible background comes from $e^- p \to e^- b \bar{b}j $ via photo production ($\gamma^* p \to b \bar{b}j$). In this process, 
the scattered 
electrons are all very close to the beam pipe which do not get registered in the detector giving rise to missing energy signature.

We have generated the signal and background events at the parton level using MG5 with the following selection cuts. 
\begin{align*}
	p_T(j) > 10~{\rm GeV}; ~ \Delta R(b,b), \Delta R(b,j) > 0.4  
\end{align*}

We use the energy smearing function given by, 
\begin{eqnarray}
\frac{\sigma_E}{E} = a/\sqrt{E} \oplus b
\end{eqnarray}
to take into account the detector effects. We choose 
$a=0.6,~b=0.04$ for parton jets, and $a=0.12,~b=0.02$ for electron~\cite{LHeC:2020van}.
After taking into account the smearing effects, we apply $p_T(j),~p_T(b) > 30$ GeV cuts. 
To take care of the photo production background, we apply $\slashed{E_T}$ cut of 25 GeV. 
As in Ref.~\cite{Han:2009pe}, we find that 
$\slashed{E_T}$ cut is indeed very effective in suppressing the photo production background.
The two jet background events are ordered in rapidity and we apply a veto on the less forward jet with $p_T > 30$ GeV.  

Next, we impose  $ | M(b, \bar{b})-M_H| < 15$ GeV cut 
which is effective on all background processes. At this stage, $ e^- p \to \nu_e {\bar t} b, {\bar t} \to {\bar b} j j$ is the only relevant 
background. 
In order to improve the significance of the signal over backgrounds, we demand that the jet is in the forward region {\it i.e.} 
$1 < |\eta_j| < 5.0$. This reduces the irreducible background considerably.  
Finally, we apply $M(H,j)>250$ GeV cut which further reduces the background.
Many of these cuts are motivated by the VBF studies performed for the LHC.
In our analysis, we use a 60\% tagging rate for $b$ jets. We note that the irreducible 
backgrounds $\nu_e jjj$ are negligible after considering $c$ jet, and light jet mistagging rates as 0.1, and 0.01 respectively. 
In table~\ref{tab:hww_cuts}, we summarize the effect of various cuts on signal and dominant background processes.

\newcolumntype{M}[1]{>{\centering\arraybackslash}m{#1}}
\renewcommand{\arraystretch}{2.5}
\begin{table}[htp]
	\centering
	\begin{tabular}{M{2cm}|M{2cm}|M{2cm}|M{2cm}|M{2.5cm}|M{2cm}}
		\hline
		\hline
		&Generation cuts after smearing
		&$p_T(j)>$ 30 GeV, $p_T(b)>$ 30 GeV, $\cancel{E}_T > 25 $GeV & $|M_{b\bar{b}}-m_H|$ $<$ 15 GeV  & $1 < \eta_j < 5.0 $, $-1 < \eta_b < 4.0$ & $M_{Hj}>$ 250 GeV\\
		\hline
		\hline
		SM signal  &  3011 & 1315 & 1296 & 1251 &  819\\
		\hline
		$e^- p \rightarrow \nu_e b \bar{b} j$ & 18883  & 1877  & 83 & 60 & 30\\
		\hline
		$e^- p \rightarrow \nu_e \bar{t} b$, $\bar{t} \rightarrow \bar{b} j j$ & 10985  & 1597  & 326 & 152 & 38 \\
		\hline
		S/B & 0.1 & 0.4 & 3.2 & 5.9 & 12.0 \\
		\hline
		\hline
	\end{tabular}
	\caption{SM signal and background events (at $\mathscr{L}=100$ $fb^{-1}$) with selection cuts for the {\tt CC} process. Signal to background ratio (S/B) is given in the last row. A tagging efficiency of 60\% for $b$ quark is assumed.} 
	\label{tab:hww_cuts}
\end{table}

In the case of the {\tt NC} process, all the final state particles can be seen in the detector. The dominant backgrounds mimicking $e^- H(b \bar{b}) j$ 
final state include (i) the irreducible background $e^- p \to e^- b \bar{b} j$ and (ii) reducible background $e^-p \to e^- b \bar{b} jj$. In this 
case, the generation cuts on particles other than $e^-$ are as before. On the $e^-$ following generation cuts are applied.
\begin{align*}
	p_T(e) > 10~{\rm GeV};~ \Delta R(e,b), \Delta R(e,j) > 0.4
\end{align*}
The signal significance is very poor due to a very large background at this stage.
{Selection cuts $p_T(e) >$ 20 GeV, $p_T(j) >$ 30 GeV and $p_T(b) >$ 30 GeV reduce dominant backgrounds almost 99$\%$ at the cost of losing half of the signal events. These cuts are very effective on the $2j$ background.
Further, we impose invariant mass cut $ | M(b, \bar{b})-M_H| < 15$ GeV which reduces the irreducible background significantly. Pseudo-rapidity cuts as given in the table \ref{tab:hzz_cuts} reduce backgrounds and improve the S/B ratio. The final cut of $M(H,j) >$ 300 GeV brings down the total background to 16$\%$ of the background at the previous stage with an increased S/B ratio of 0.41.} A cut-flow chart for the {\tt NC} process is presented in table~\ref{tab:hzz_cuts}.

The cuts discussed for the {\tt CC} and {\tt NC} standard model processes are kept fixed for non-zero values of the BSM parameters. We assume that cut efficiencies do not change substantially for reasonable values of BSM parameters. 

\newcolumntype{M}[1]{>{\centering\arraybackslash}m{#1}}
\renewcommand{\arraystretch}{2.5}
\begin{table}[htp]
	\centering
	\begin{tabular}{M{2cm}|M{2cm}|M{2cm}|M{2cm}|M{2.5cm}|M{2cm}}
		\hline
		\hline
		&Generation level cuts
		& $p_T(e)>$ 20 GeV, $p_T(j)>$ 30 GeV, $p_T(b)>$ 30 GeV &  $|M_{b\bar{b}}-m_H|$ $\leq$ 15 GeV &  $ |\eta_e| < 2.5$, $2 < \eta_j < 5$, $0.5 < \eta_b < 3$ & $M_{Hj}>$ 300 GeV\\
		\hline
		\hline
		SM signal  & 534 & 274 & 270 & 161 & 76\\
		\hline
		$e^- p \rightarrow e^- b \bar{b} j$ & 2.75 $\times$ $10^6$ & 1.3 $\times$ $10^4$ & 2425 & 835 & 161\\
		\hline
		$e^- p \rightarrow e^- b \bar{b} j j$ & 6.3 $\times$ $10^5$ & 4218 & 789 & 336 & 24\\
		\hline
		S/B &  0.02 $\times 10^{-2}$  & 0.02 & 0.08 & 0.14 & 0.41\\
		\hline
		\hline
	\end{tabular}
	\caption{SM signal and background events (at $\mathscr{L}=100$ $fb^{-1}$) with selection cuts for the {\tt NC} process. Signal to background ratio (S/B) is given in the last row. We have assumed $b$ tagging efficiency of 60\%.}
	\label{tab:hzz_cuts}
\end{table}

\section{Projected constraints on BSM parameters}
\label{sec:results}

In order to estimate constraints on anomalous couplings, we perform $\chi^2$ analysis assuming the standard model hypothesis. 
The analysis is done first at the inclusive level and then using $\Delta \phi$ 
distribution for each process. For a given BSM parameter $c_i$, the $\chi^2$ function 
is given by, 
\begin{align}
\chi^2(c_{i})=\sum_{j=1}^{n}\left( \frac{N_j^{\rm BSM}(c_{i})-N_j^{\rm SM}}{\Delta N_j} \right)^2,
\end{align}
where $N_j^{\rm SM}$ and $N_j^{\rm BSM}$ are the numbers of SM and BSM events in the $j^{\rm th}$ bin of the $\Delta \phi$ distribution 
after applying all the cuts and efficiencies discussed in the previous section.  
The uncertainty in the $j^{\rm th}$ bin, $\Delta N_j$ includes both statistical and systematic uncertainties. 
It is given by, 
\begin{align}
	\Delta N_j^{\rm SM+Bkg} = \sqrt{N_j^{\rm SM+Bkg} \left( 1+\delta_{sys}^2 N_j^{\rm SM+Bkg} \right) }.
\end{align}

For our analysis, presented in the next subsections, we choose 5\% systematic uncertainty uniformly in all the bins. Due to limited statistics at low luminosity, we restrict our analysis to two bins of the $\Delta \phi$ distribution. We choose these two bins symmetrically at about $\Delta \phi=\pi/2$. Table~\ref{tab:xsection_coeff} can be used to obtain the cross section as a function of $c_i$  in each bin after applying the final analysis cuts.  
In the following, we will refer to the analysis based on the total cross section as 1 bin analysis, while the analysis based on $\Delta \phi$ will be referred to as 2 bin analysis.

\begin{table}[ht]
	\centering
	\begin{tabular}{M{2cm}|M{1.5cm}|M{1.5cm}|M{1.5cm}|M{1.5cm}|M{1.5cm}|M{1.5cm}}
		\hline
		\hline
		& \multicolumn{3}{c|}{First Bin} & \multicolumn{3}{c}{Second Bin} \\
		\hline
		\hline
		Couplings  &$X_{ii}$ ($pb$)
		& $X_{i}$ ($pb$) & $X_{SM}~(pb)$ & $X_{ii}$ ($pb$) & $X_{i}$ ($pb$) & $X_{SM}~(pb)$\\
		\hline
		$\lambda_{1W}$ & 0.0140 & -0.0052 &        & 0.0271 & 0.0123 & \\
		
		$\lambda_{2W}$ & 0.0737 & -0.0218 & 0.0033 & 0.1781 & -0.0388 & 0.0049  \\
		
		$\tilde{\lambda}_W$ & 0.0159 & 0 & & 0.0176 & 0 & \\
		\hline
		
		$\lambda_{1Z}$ & 0.0008 & -0.0005 &        & 0.0012 & 0.0008 & \\
		
		$\lambda_{2Z}$ & 0.0067 & -0.0022 & 0.0003 & 0.0109 & -0.0029 & 0.0004 \\
		
		$\tilde{\lambda}_Z$ & 0.0009 & 0 & & 0.0009 & 0 & \\
		\hline
		\hline
	\end{tabular}
	\caption{Coefficients of BSM parameters given in Eq.~\ref{xsection} for $\Delta \phi$ distribution 
	after applying all the cuts given in tables~\ref{tab:hww_cuts} and~\ref{tab:hzz_cuts}. Similar coefficients at the total cross section level can be calculated from the two bin information. These numbers are used in the one parameter analysis of section~\ref{sec:1p}. } 
\label{tab:xsection_coeff}
\end{table}

\begin{figure}
	\centering
	\subfloat[]{\includegraphics[width=.50 \textwidth]{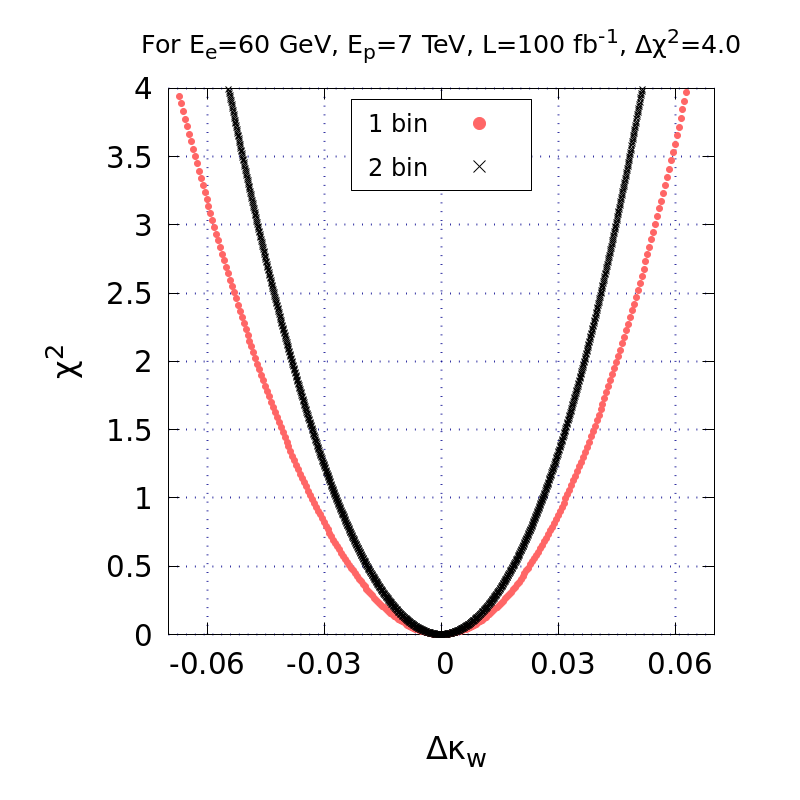}}
	\subfloat[]{\includegraphics[width=.50 \textwidth]{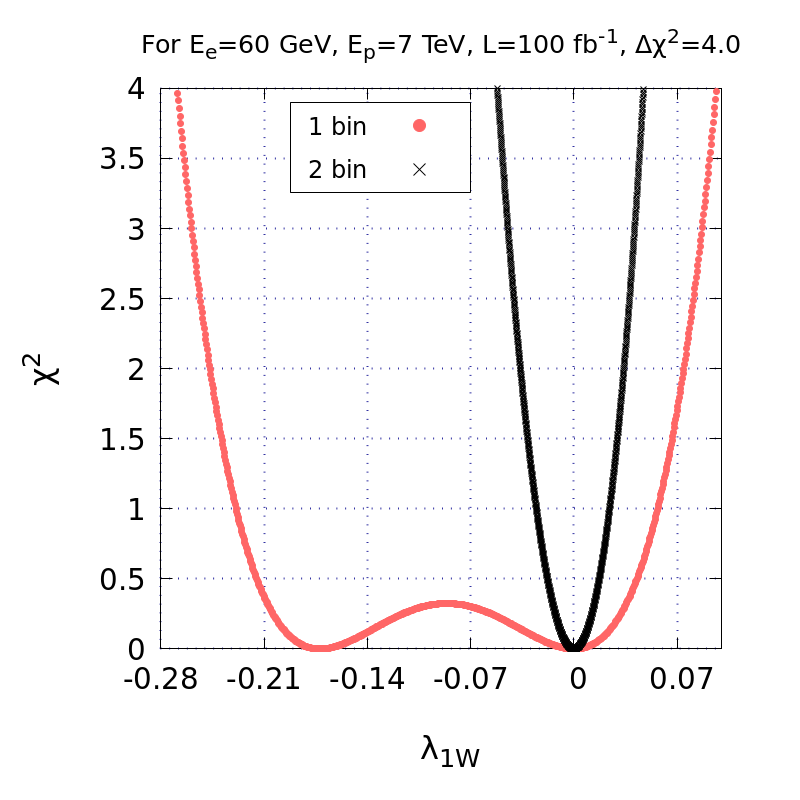}}\\
	\subfloat[]{\includegraphics[width=.50 \textwidth]{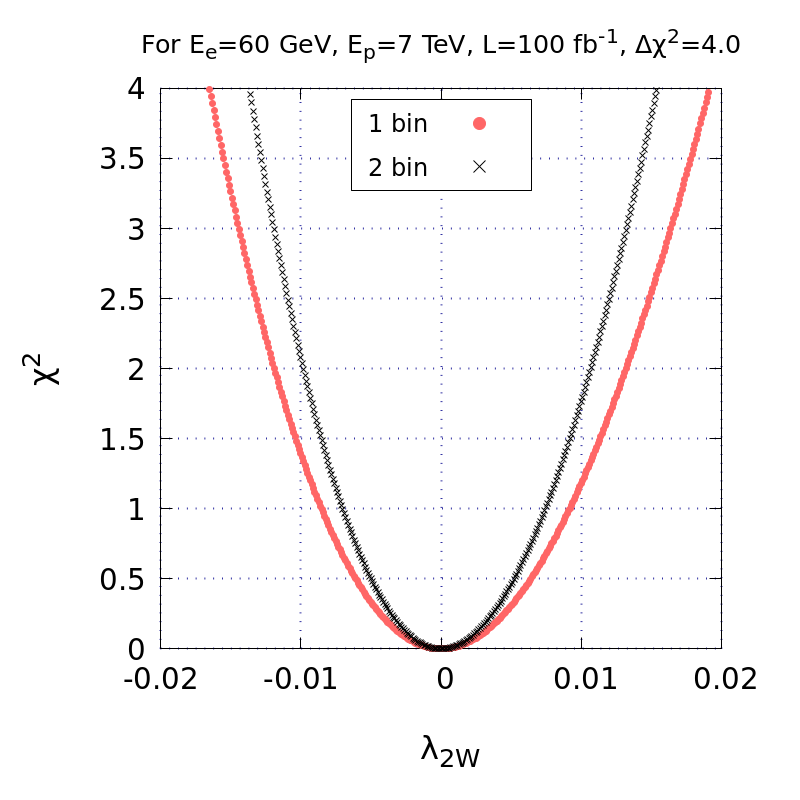}}
	\subfloat[]{\includegraphics[width=.50 \textwidth]{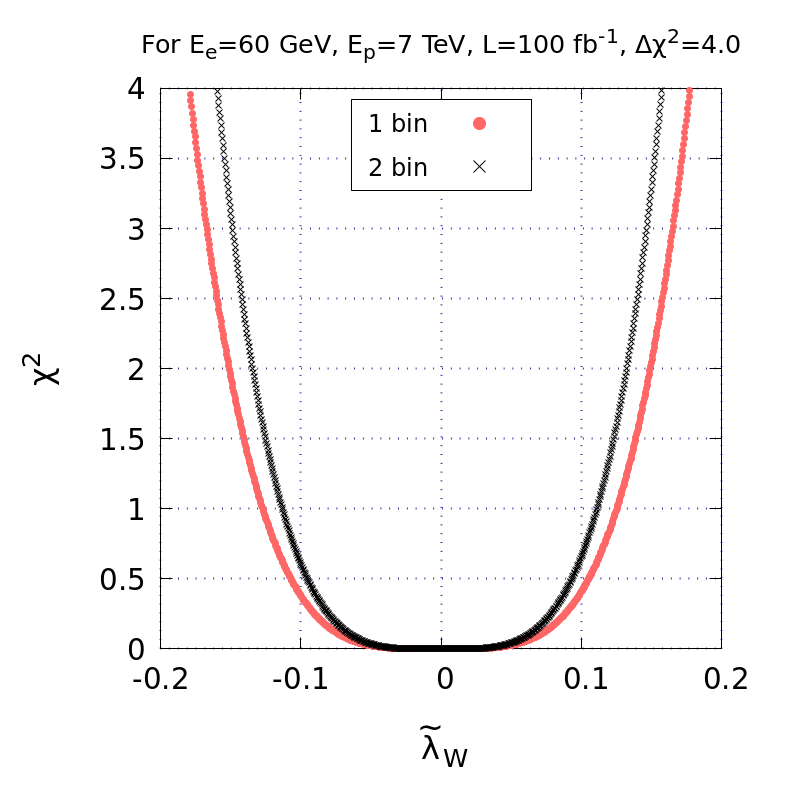}}
	\caption{$\chi^2$ distribution for 1 bin (red curve) and 2 bins (black curve) of $\Delta \phi$ distribution in the {\tt CC} process. In (a), we have defined $\Delta \kappa_W = 1-\kappa_W$.}
	\label{fig:chisq_wwh_1p}
\end{figure}

\begin{figure}
        \centering
	\subfloat[]{\includegraphics[width=.50 \textwidth]{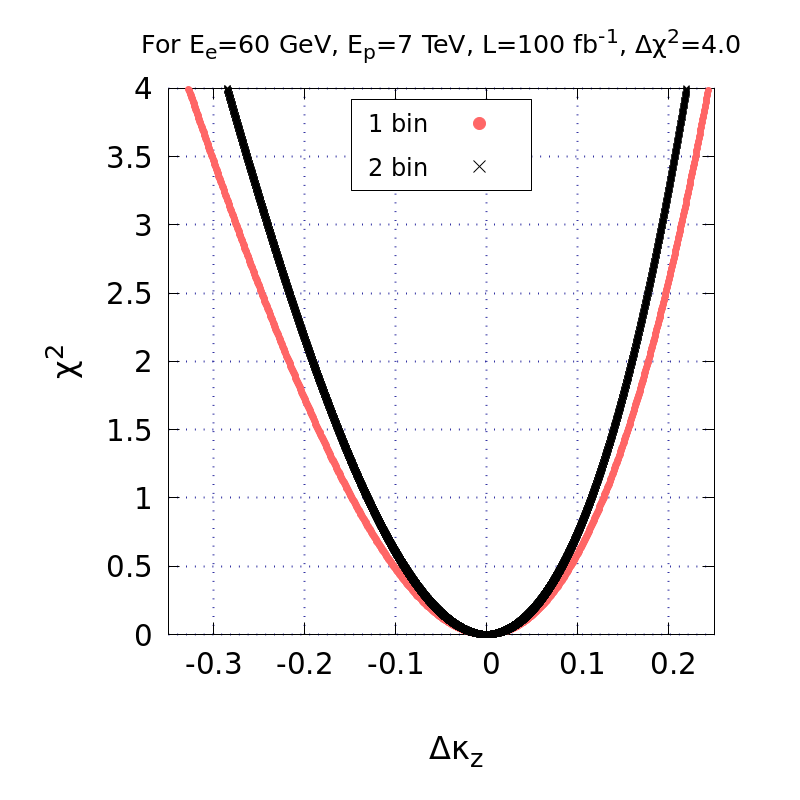}}
	\subfloat[]{\includegraphics[width=.50 \textwidth]{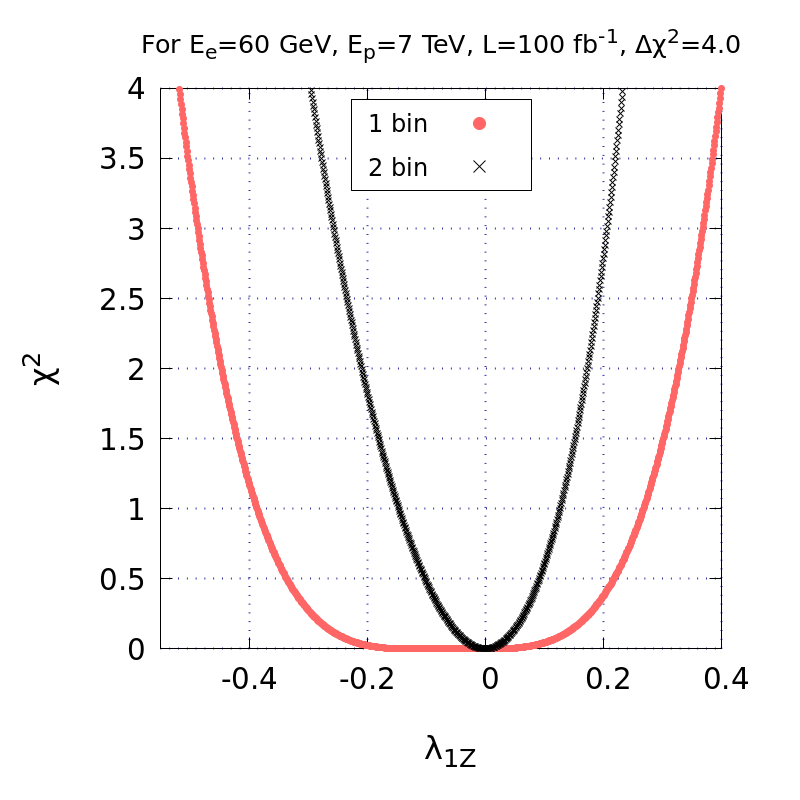}}\\
	\subfloat[]{\includegraphics[width=.50 \textwidth]{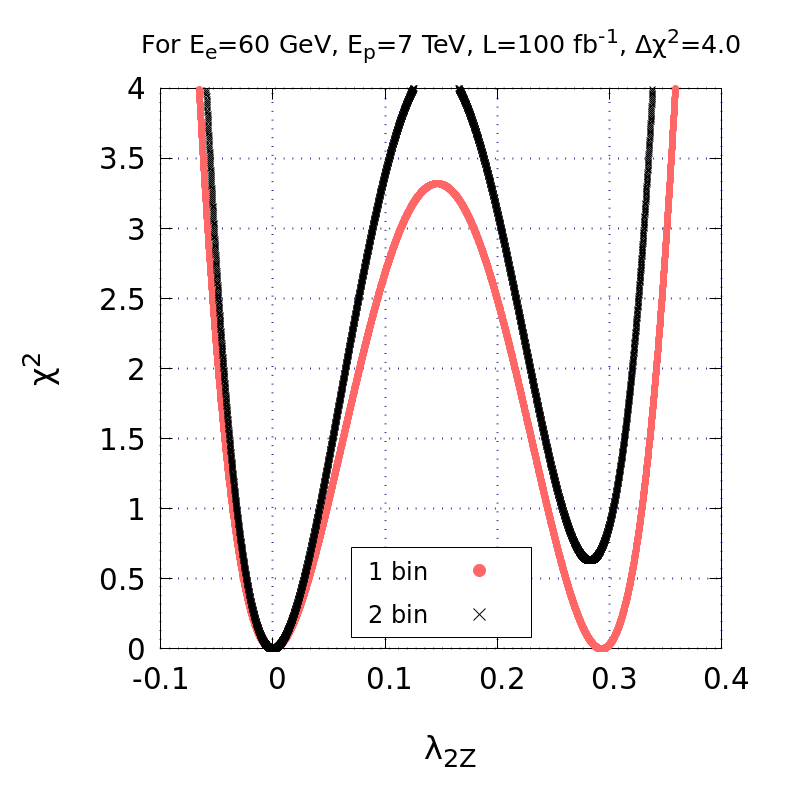}}
	\subfloat[]{\includegraphics[width=.50 \textwidth]{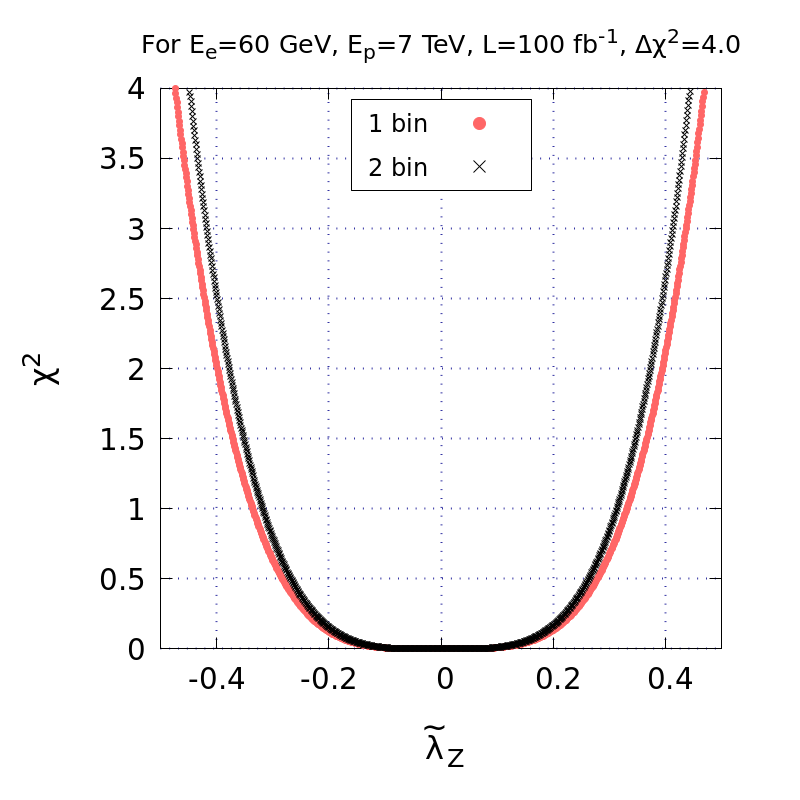}}
	\caption{$\chi^2$ distribution for 1 bin (red curve) and 2 bins (black curve) of $\Delta \phi$ distribution in the {\tt NC} process. In (a), we have defined $\Delta \kappa_Z = 1-\kappa_Z$.}
        \label{fig:chisq_zzh_1p}
\end{figure}

\subsection{One parameter analysis}
\label{sec:1p}
First, we consider the case in which only one of the four parameters is taken non-zero at a time. The results of one parameter analysis 
for an integrated luminosity of 100 $fb^{-1}$ are shown in figures~\ref{fig:chisq_wwh_1p} and~\ref{fig:chisq_zzh_1p}. 
We find that the 2 bin analysis improves the constraints on all the BSM parameters. From figures \ref{fig:chisq_wwh_1p}(a) and \ref{fig:chisq_zzh_1p}(a), we note that 
the constraints on $\kappa_W$ improve by 26\% while constraints on $\kappa_Z$ improve by 12\% in switching from 1 bin to 2 bin analysis. At $95\%$ C.L., the constraints on $\kappa_W$ and $\kappa_Z$ are [0.94,1.05] and [0.72,1.22], respectively. The fit based on Run-II data (35.9 $fb^{-1}$) of 13 TeV LHC gives $\kappa_W \in [0.76,1.34]$ and $\kappa_Z \in [0.75,1.21]$ at $2\sigma$~\cite{CMS:2018uag}.  

In the {\tt CC} process, the 2 bin analysis leads to significant improvement in constraints on $\lambda_{1W}$ and $\lambda_{2W}$. 
For example, at 95\% C.L. the allowed region for $\lambda_{1W}$ in figure~\ref{fig:chisq_wwh_1p}(b) changes from [-0.26, 0.1] to [-0.05, 0.05] when we change the analysis from 
1 bin to 2 bins.
Note that the two bin analysis is able to break the degeneracy in $\chi^2$ minimum for $\lambda_{1W}$ which appears at $\lambda_{1W}=0$ and 
$\lambda_{1W}=-0.17$. A similar feature is 
observed for $\lambda_{2W}$ in which case the 2nd minimum in $\chi^2$ appears at $\lambda_{2W}=-0.24$. The separation between the two minima is sensitive to the size of the linear term which is much larger in the case of $\lambda_{2W}$ than in $\lambda_{1W}$ (see table~\ref{tab:xsection_coeff}).
Since the overall constraints on $\lambda_{2W}$ are much tighter, in figure \ref{fig:chisq_wwh_1p}(c) we have shown the constraints only 
about $\lambda_{2W}=0$, the minima chosen by the 2 bin analysis. The allowed regions for $\lambda_{2W}$ are [-0.016, 0.018] and [-0.013, 0.015] for 1 bin and 2 bin analysis, 
respectively. 
Since the cross section depends on ${\tilde \lambda}_W$ quadratically, the two bin analysis leads to only a
slight improvement in the constraints as visible in figure \ref{fig:chisq_wwh_1p}(d). In going from 1 bin to 2 bin analysis, the allowed region changes 
from [-0.18, 0.18] to [-0.16, 0.16], accounting for a 10\% improvement. 
We note that these limits are consistent with the limits obtained on $\lambda_{1W}$ and ${\tilde \lambda}_{W}$ in Ref.~\cite{Biswal:2012mp} with $E_e=140$ GeV and $E_p=6.5$ TeV. The parameters of $HWW$ have been constrained by studying the double Higgs production at the FCC-he ($E_e=60$ GeV, $E_p$= 50 TeV), however, those limits are much weaker than ours~\cite{Kumar:2015kca}.

In the {\tt NC} process, constraints on BSM parameters are less stringent as compared to {\tt CC} process since {\tt NC} process is background dominated over signal events. Among all BSM parameters, $\lambda_{1Z}$ is constrained the most as we go from 1 bin to 2 bin analysis. The 2 bin analysis improves the constraint by 41\% for this parameter.
Unlike in the case of $\lambda_{1W}$, the degenerate minima in the 1 bin analysis of $\lambda_{1Z}$ are not well separated leading to a flat region at the bottom of the plot. The flatness is removed when using the 2 bin information. In the case of $\lambda_{2Z}$, the 2 bin analysis breaks the degeneracy, however, a significant region about the second minimum at $\lambda_{2Z}=0.28$ is still allowed.  
The allowed region for $\lambda_{2Z}$ is [-0.06, 0.36] for 1 bin analysis while [-0.06,0.13] $\cup$ [0.17, 0.34] is for 2 bin analysis. Only at a very large luminosity, the second region of the 2 bin analysis can be ruled out. 
We find that there are no significant changes in the constraints on $\widetilde{\lambda}_Z$ when choosing 2 bin over 1 bin analysis. It is due to the same size of the coefficients of $\widetilde{\lambda}_Z^2$ term in both the bins as given in table \ref{tab:xsection_coeff}. The slight improvement in constraints is mainly due to the difference in the errors in the two cases.

 We have found that the above constraints improve with an increase in the 
beam energies of the electron and the proton as expected. However, the effect of increasing electron energy on constraints 
is more than increasing the proton beam energy. We have also checked that there is no significant improvement in constraints from the {\tt NC} 
process if we use -80\% polarized electron beam. In the {\tt CC} process,  $\Delta \phi$ distribution increases 1.8 times in each bin for -80\% polarized electron beam which leads to about 7-10\% improvements in constraints on $HWW$ parameters.

\begin{figure}
	\centering
	\subfloat[]{\includegraphics[width=.45 \textwidth]{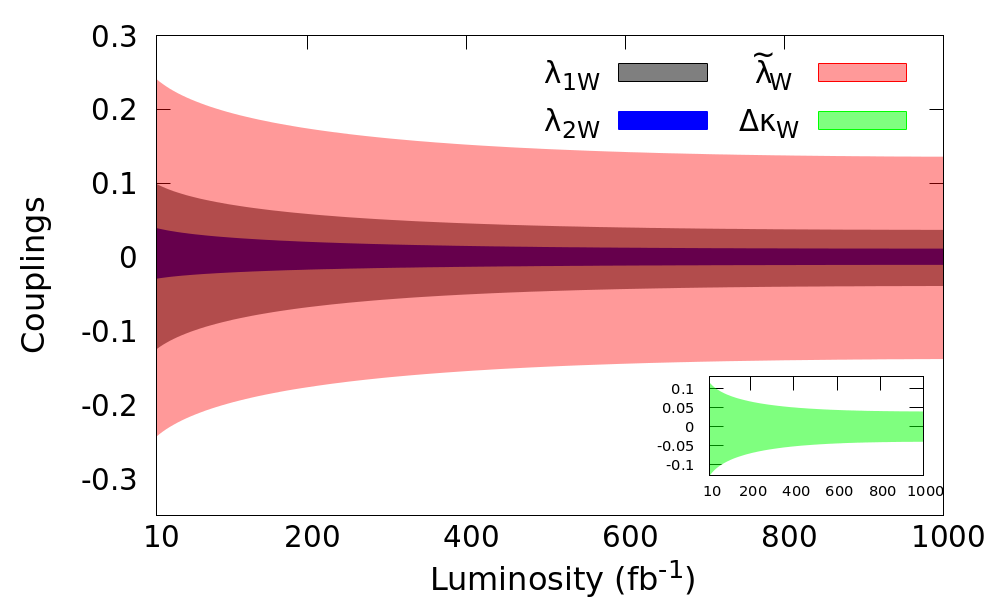}\label{fig:L_vs_l_wwh}} 
	\subfloat[]{\includegraphics[width=.45 \textwidth]{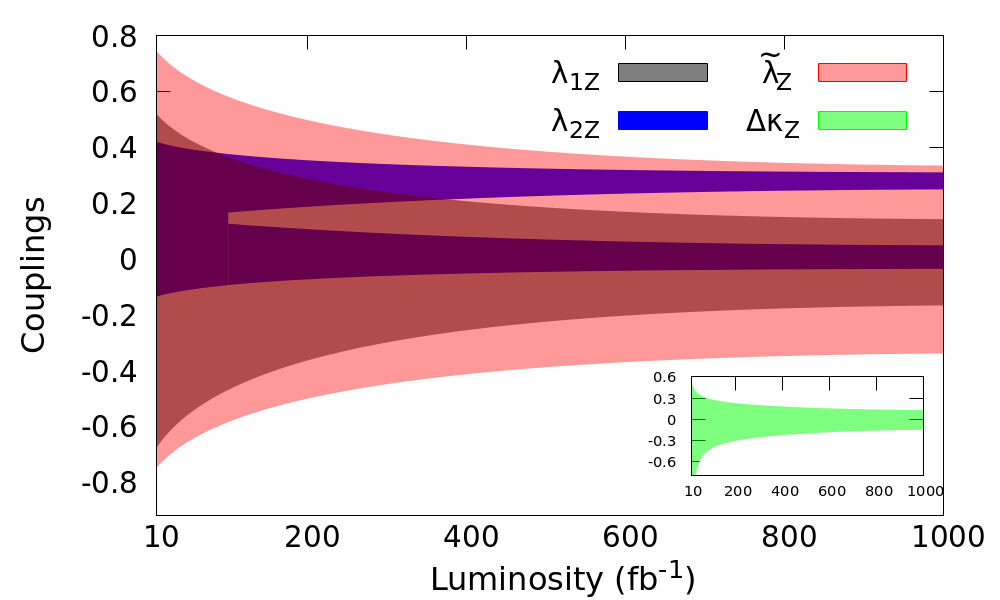}\label{fig:L_vs_l_zzh}}
	\caption{Luminosity vs BSM coupling parameters (a) {\tt CC} process (b) {\tt NC} process. }
	\label{fig:L_vs_l_coup}
\end{figure}

In figure~\ref{fig:L_vs_l_coup}, we present one parameter constraints at 95\% C.L. as a function of luminosity. We have varied the luminosity 
from 10 $fb^{-1}$ to 1000 $fb^{-1}$. All BSM parameters decrease with 
increasing luminosity and follow the same trend. Changing the luminosity from 10 $fb^{-1}$ to 1000 $fb^{-1}$ improves the constraints by 67\%, 40\%, 78\%, and 42\% on $\kappa_W$, $\lambda_{1W}$, $\lambda_{2W}$, and $\widetilde{\lambda}_W$, respectively. In the {\tt NC} process, unlike $\kappa_{Z}, \lambda_{1Z}$, and $\widetilde{\lambda}_Z$ parameters, $\lambda_{2Z}$ has two branches of the 
allowed region. The allowed regions for $\kappa_Z$, $\lambda_{1Z}$, $\lambda_{2Z}$, and $\widetilde{\lambda}_Z$ shrink by 91\%, 74\%, 73\%, and 55\%, respectively as the luminosity is changed from 10 $fb^{-1}$ to 1000 $fb^{-1}$. Projected constraints for $\mathscr{L}=1000~ fb^{-1}$ are summarized in table~\ref{tab:constraints-1000}. For reference we mention that 
at 2$\sigma$, the expected reach 
on $\kappa_W$ and $\kappa_Z$ are 3.4\% (1.6\%)  and 3.0\% (0.8\%), respectively at HL-LHC (CLIC: 350 GeV, 
1 $ab^{-1}$)~\cite{Cepeda:2019klc,Robson:2018zje}. The current limits on $CP$-odd parameters obtained using the Run-II LHC data are 
${\tilde \lambda}_Z = \frac{1}{2} {\tilde \lambda}_W \in [-0.21,0.15]$ at $2\sigma$~\cite{ATLAS:2020evk}. The projections at HL-LHC for 
$\lambda_{1Z}, \lambda_{2Z}$, and ${\tilde \lambda}_{Z}$ are about 1\%, 0.7\%, and 12\%, respectively~\cite{Cepeda:2019klc, Boselli:2017pef}.

\begin{table}
	\centering
\begin{tabular}{|c||c| c|c|c|}
	\hline
	BSM parameter & $\kappa_W$ & $\lambda_{1W}$ & $\lambda_{2W}$ & $\widetilde{\lambda}_{W}$ \\
	& $\kappa_Z$ & $\lambda_{1Z}$ &  $\lambda_{2Z}$ & $\widetilde{\lambda}_{Z}$ \\
	\hline
Constraint & [0.96, 1.04] & [-0.04, 0.04] & [-0.01, 0.01] & [-0.14, 0.14]\\
& [0.85, 1.13] & [-0.17, 0.14]& [-0.04, 0.05] $\cup$ [0.25, 0.31] & [-0.34, 0.34] \\
	\hline
\end{tabular}
	\caption{Projected constraints on BSM parameters with 1000 $fb^{-1}$ integrated luminosity at 95\% C.L.}
	\label{tab:constraints-1000}
\end{table}

\subsection{Two parameter analysis}
\label{sec:2p}
\begin{figure}[htp]
	\centering
	\subfloat[]{\includegraphics[width=.33
	\textwidth]{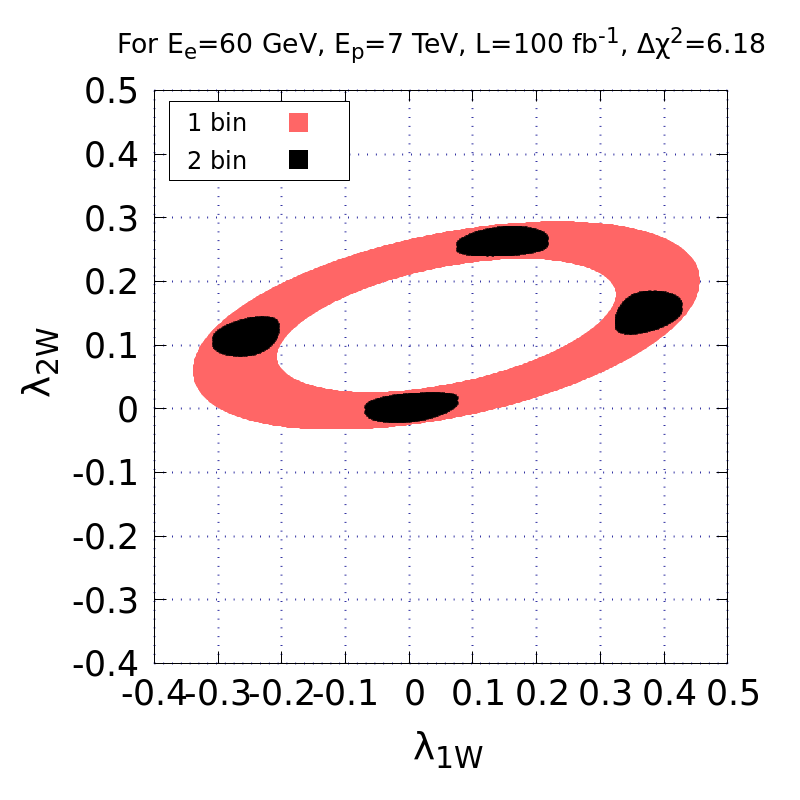}}
	\subfloat[]{\includegraphics[width=.33 \textwidth]{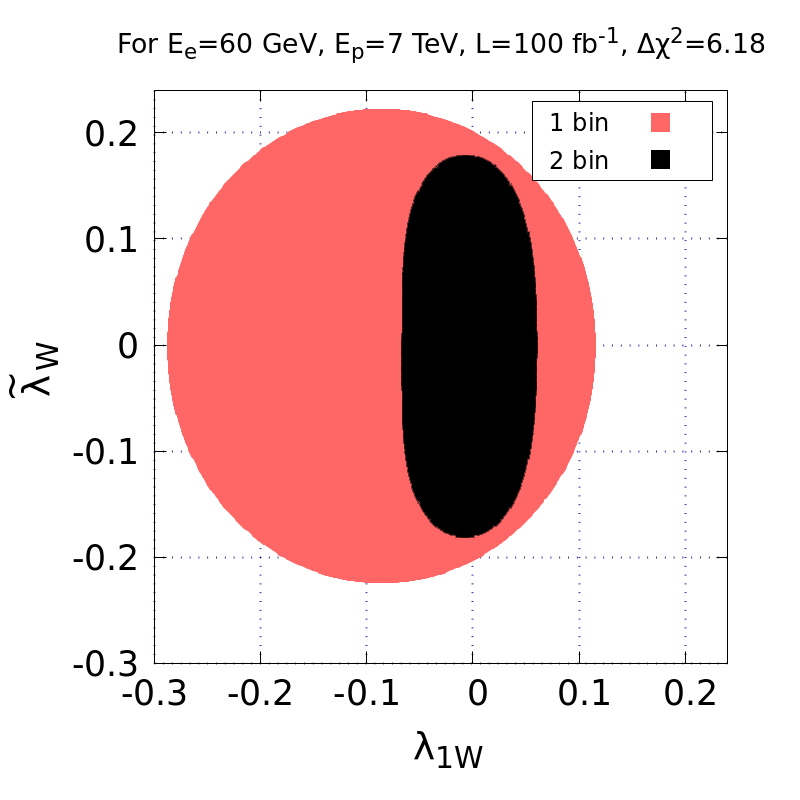}}
	\subfloat[]{\includegraphics[width=.33 \textwidth]{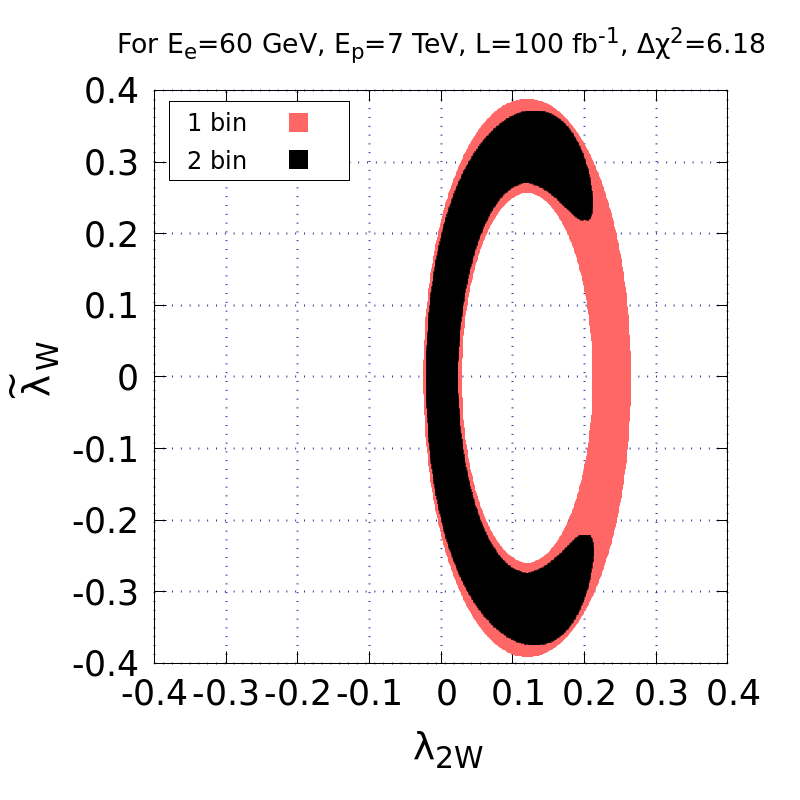}}
	\caption{Two-dimensional parameter space for $HWW$ BSM parameters at $\mathscr{L}=100~fb^{-1}$ using cross section (red) and 
	$\Delta \phi$ distribution (black) as observables.}
	\label{fig:2param_hww}
\end{figure}

\begin{figure}[htp]
	\centering
	\subfloat[]{\includegraphics[width=.33 \textwidth]{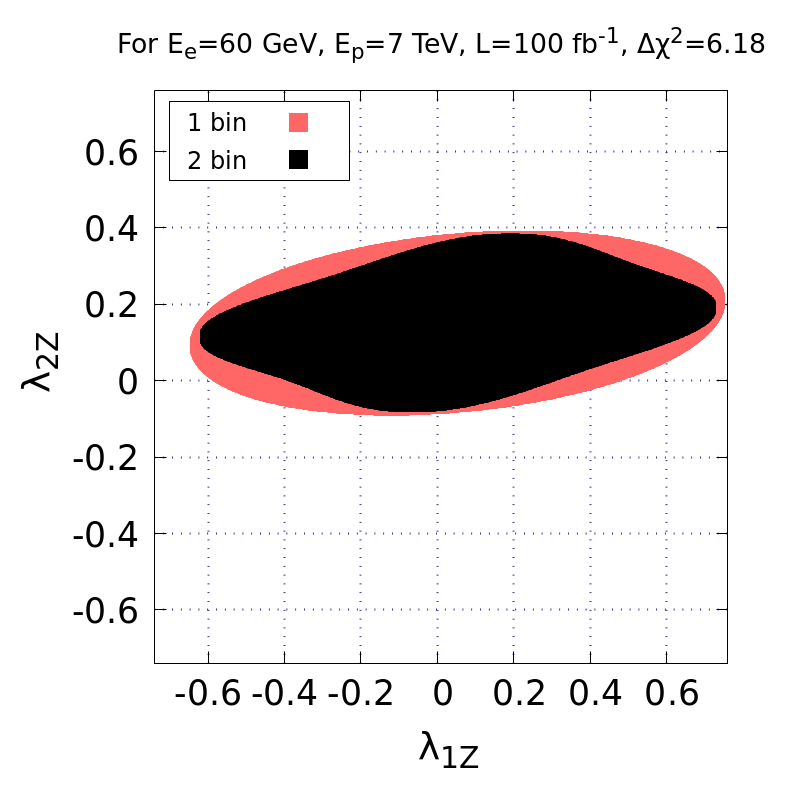}}
	\subfloat[]{\includegraphics[width=.33 \textwidth]{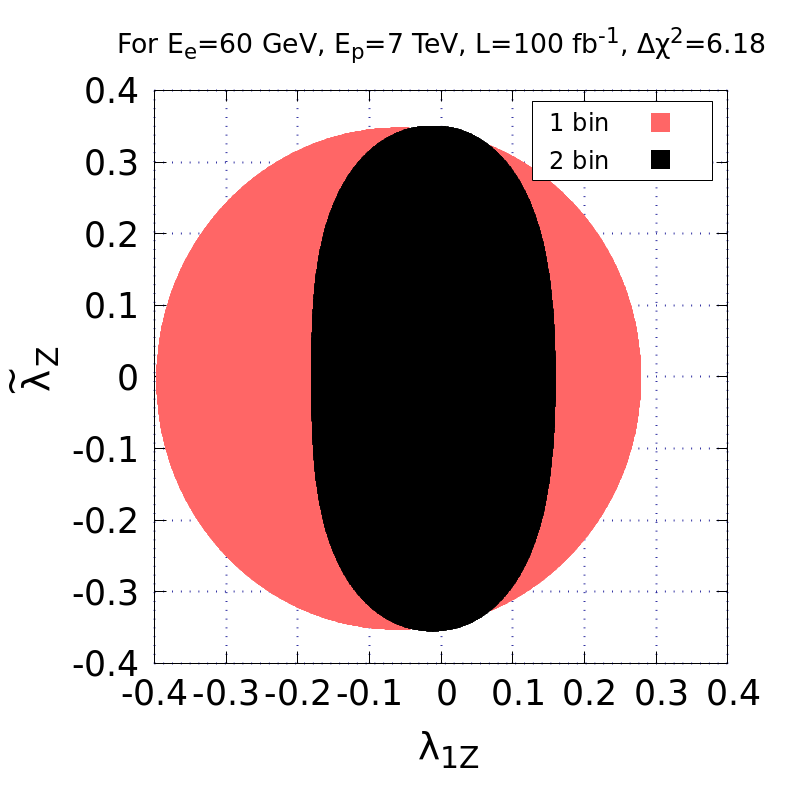}}
	\subfloat[]{\includegraphics[width=.33 \textwidth]{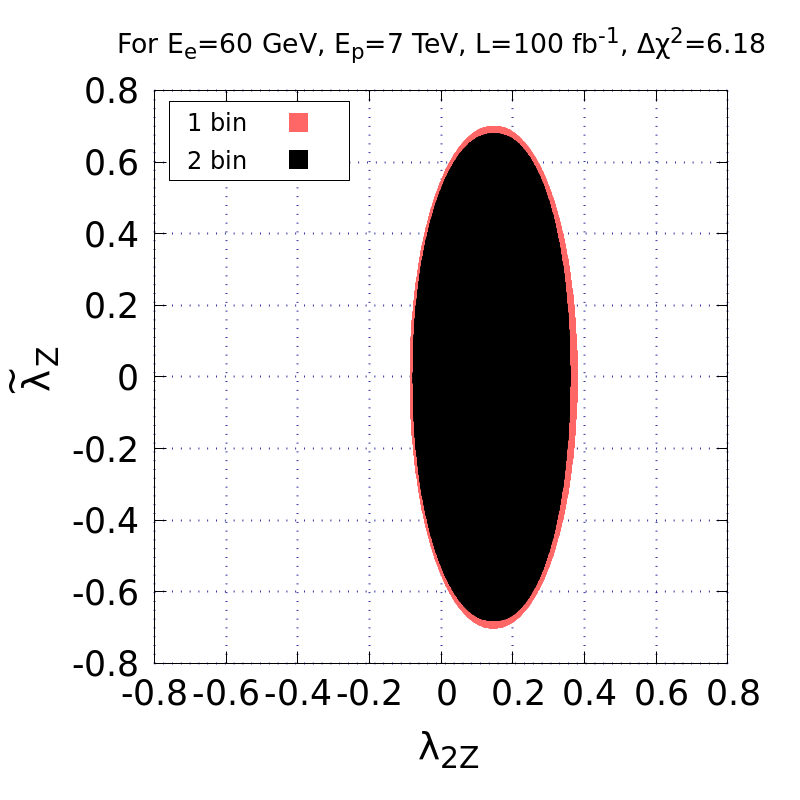}}
	\caption{Two-dimensional parameter space for $HZZ$ BSM parameters at $\mathscr{L}=100~fb^{-1}$ using cross section (red) and 
	$\Delta \phi$ distribution (black) as observables. }
	\label{fig:2param_hzz}
\end{figure}

The results of one parameter analysis discussed above provide the most conservative bound on each BSM parameter. 
These bounds are useful in a scenario when two out of three nontrivial BSM parameters can be constrained severely 
using some other observables.    
In a more general scenario, one would like to know how constraints  
on a given BSM parameter change in presence of other BSM parameters.
For that, we consider the case in which two of the three non-trivial BSM couplings are taken non-zero at a time. 
In figures~\ref{fig:2param_hww} and ~\ref{fig:2param_hzz}, two parameter spaces consistent with the SM hypothesis at 
95\% CL are shown for the {\tt CC} and {\tt NC} processes, respectively. An integrated luminosity of 100 $fb^{-1}$ has been assumed. 
The plots are obtained by taking $\Delta \chi^2$=6.18 around minima. This value corresponds to 95\% C.L. for 2 parameter case~\cite{Workman:2022}. 

In each figure, we display the effect of using differential distribution in the $\chi^2$ analysis. Two parameter analysis results are consistent with one parameter results. We note that the efficacy of 2 bin analysis over 1 bin analysis is stronger on the parameters of the {\tt CC} process as compared to those of the {\tt NC} process. Since $\lambda_{1V}$ and $\lambda_{2V}$ are $CP$-even parameters, the constraints on them are correlated, that is, the coefficient of cross-term in the observable is sufficiently large.

Contour plots for 1 bin analysis in the planes of ($\lambda_{1W}, \lambda_{2W}$) and ($\lambda_{2W}, \widetilde{\lambda}_W$) are ring-shaped. 
This is related to the fact that the 1 parameter fit for $\lambda_{2W}$  using the total cross section information results in two well-separated allowed regions. This happens for some non-zero values of $\lambda_{1W}$ and ${\tilde \lambda}_{W}$ as well. The effect of the 2 bin analysis is most prominent in the ($\lambda_{1W},\lambda_{2W}$) plane
where the allowed region reduces to four disconnected regions in figure~\ref{fig:2param_hww}(a). Depending on the value of $\lambda_{1W}$, large values of $\lambda_{2W}$ such as 0.25 are allowed. It is clear from figure  \ref{fig:2param_hww}(c) that this value, however, is not compatible with any value of $\widetilde{\lambda}_W$ when $\lambda_{1W}=0$. The region constrained by 1 bin analysis, in the plane of ($\lambda_{1W}, \widetilde{\lambda}_W$) is almost a circular disc which, when using 2 bin analysis, shrinks into an elliptical region with the major axis about $\lambda_{1W}=0$.

In the {\tt NC} process, parameter space in the plane of ($\lambda_{1Z}, \widetilde{\lambda}_Z$) shows significant improvement when 2 bin analysis is used. The constraints in other cases change only marginally. Substantially large data would be required in order to constrain the two parameter regions of the {\tt NC} process.

\section{Conclusions}
\label{sec:concl}
We have studied the {\tt CC} and {\tt NC} processes for single Higgs production at an $e^- p$ collider with $E_e=60$ GeV and $E_p=7000$ GeV. The effects 
of the most general $HVV$ vertices, which are relevant to these processes, on the total cross sections and the $\Delta \phi$ 
distributions have been studied. We find that using $\Delta \phi$ distribution in the analysis over the total cross section 
leads to stronger bounds on the BSM parameters. Constraints obtained for $CP$-even parameters are tighter than those on $CP$-odd parameters. 

We have obtained the constraints in two scenarios; ($i$) only one BSM parameter is non-zero at a time and ($ii$) two BSM parameters are non-zero at a time. Assuming the SM hypothesis, the one parameter constraints lie in the range of 1-15\% for $HWW$ parameters, and 5-34\% for $HZZ$ parameters with 1000 $fb^{-1}$ data at 95\% C.L. These constraints change considerably when some other BSM parameters are present. In the case of two parameter analysis, the $\Delta \phi$ distribution is most effective in constraining $(\lambda_{1W},\lambda_{2W}), (\lambda_{1W},{\tilde \lambda}_{W}),  (\lambda_{2W},{\tilde \lambda}_{W})$ and ($\lambda_{1Z},{\tilde \lambda}_{Z}$) parameter regions. The projected constraints on $HWW$ parameters $\kappa_W$ and ${\tilde \lambda}_{W}$, from our 
analysis of the {\tt CC} process, are compatible with the expected reach at the future colliders HL-LHC and CLIC. However, the constraints on 
$HZZ$ parameters from the study of the {\tt NC} process are weaker than the projections at future colliders.  

\acknowledgments
{We would like to acknowledge fruitful discussions with Rafiqul Rahaman on the signal-background analysis. We thank Mukesh Kumar for various discussions and his participation in the initial stages of the project. Special thanks to Olivier Mattelaer for his help with the MG5 package. PS would like to acknowledge financial support from IISER Mohali for this work.}

\pagebreak

\bibliographystyle{JHEP}
\bibliography{bibref}

\providecommand{\href}[2]{#2}\begingroup\raggedright\begin{thebibliography}{10}

\bibitem{GLASHOW:1961579}
S.~L. Glashow, \emph{Partial-symmetries of weak interactions},
  \href{http://dx.doi.org/https://doi.org/10.1016/0029-5582(61)90469-2}{\emph{Nuclear
  Physics} {\bf 22} (1961) 579--588}.

\bibitem{Englert:13321}
F.~Englert and R.~Brout, \emph{Broken symmetry and the mass of gauge vector
  mesons}, \href{http://dx.doi.org/10.1103/PhysRevLett.13.321}{\emph{Phys. Rev.
  Lett.} {\bf 13} (Aug, 1964) 321--323}.

\bibitem{HIGGS:1964132}
P.~Higgs, \emph{Broken symmetries, massless particles and gauge fields},
  \href{http://dx.doi.org/https://doi.org/10.1016/0031-9163(64)91136-9}{\emph{Physics
  Letters} {\bf 12} (1964) 132--133}.

\bibitem{Higgs:13508}
P.~W. Higgs, \emph{Broken symmetries and the masses of gauge bosons},
  \href{http://dx.doi.org/10.1103/PhysRevLett.13.508}{\emph{Phys. Rev. Lett.}
  {\bf 13} (Oct, 1964) 508--509}.

\bibitem{Guralnik:13585}
G.~S. Guralnik, C.~R. Hagen and T.~W.~B. Kibble, \emph{Global conservation laws
  and massless particles},
  \href{http://dx.doi.org/10.1103/PhysRevLett.13.585}{\emph{Phys. Rev. Lett.}
  {\bf 13} (Nov, 1964) 585--587}.

\bibitem{Weinberg:191264}
S.~Weinberg, \emph{A model of leptons},
  \href{http://dx.doi.org/10.1103/PhysRevLett.19.1264}{\emph{Phys. Rev. Lett.}
  {\bf 19} (Nov, 1967) 1264--1266}.

\bibitem{Salam:xyz}
A.~Salam, \emph{Weak and electromagnetic interactions}, {\emph{Proceedings of
  the eighth Nobel symposium} (1968) }.

\bibitem{ATLAS:2012yve}
{\scshape ATLAS} collaboration, G.~Aad et~al., \emph{{Observation of a new
  particle in the search for the Standard Model Higgs boson with the ATLAS
  detector at the LHC}},
  \href{http://dx.doi.org/10.1016/j.physletb.2012.08.020}{\emph{Phys. Lett. B}
  {\bf 716} (2012) 1--29}, [\href{http://arxiv.org/abs/1207.7214}{{\tt
  1207.7214}}].

\bibitem{CMS:2012qbp}
{\scshape CMS} collaboration, S.~Chatrchyan et~al., \emph{{Observation of a New
  Boson at a Mass of 125 GeV with the CMS Experiment at the LHC}},
  \href{http://dx.doi.org/10.1016/j.physletb.2012.08.021}{\emph{Phys. Lett. B}
  {\bf 716} (2012) 30--61}, [\href{http://arxiv.org/abs/1207.7235}{{\tt
  1207.7235}}].

\bibitem{CMS:2013btf}
{\scshape CMS} collaboration, S.~Chatrchyan et~al., \emph{{Observation of a New
  Boson with Mass Near 125 GeV in $pp$ Collisions at $\sqrt{s}$ = 7 and 8
  TeV}}, \href{http://dx.doi.org/10.1007/JHEP06(2013)081}{\emph{JHEP} {\bf 06}
  (2013) 081}, [\href{http://arxiv.org/abs/1303.4571}{{\tt 1303.4571}}].

\bibitem{deBlas:2019rxi}
J.~de~Blas et~al., \emph{{Higgs Boson Studies at Future Particle Colliders}},
  \href{http://dx.doi.org/10.1007/JHEP01(2020)139}{\emph{JHEP} {\bf 01} (2020)
  139}, [\href{http://arxiv.org/abs/1905.03764}{{\tt 1905.03764}}].

\bibitem{ATLAS:2015xyz}
{\scshape ATLAS} collaboration, {ATLAS-CONF-2015-044}.

\bibitem{CMS:2015xyz}
{\scshape CMS} collaboration, {CMS-PAS-HIG-15-002}.

\bibitem{Cepeda:2019klc}
M.~Cepeda et~al., \emph{{Report from Working Group 2}: {Higgs Physics at the
  HL-LHC and HE-LHC}},
  \href{http://dx.doi.org/10.23731/CYRM-2019-007.221}{\emph{CERN Yellow Rep.
  Monogr.} {\bf 7} (2019) 221--584},
  [\href{http://arxiv.org/abs/1902.00134}{{\tt 1902.00134}}].

\bibitem{LHCHiggsCrossSectionWorkingGroup:2016ypw}
{\scshape LHC Higgs Cross Section Working Group} collaboration, D.~de~Florian
  et~al., \emph{{Handbook of LHC Higgs Cross Sections: 4. Deciphering the
  Nature of the Higgs Sector}},  \href{http://arxiv.org/abs/1610.07922}{{\tt
  1610.07922}}.

\bibitem{BUCHMULLER1986621}
W.~Buchmüller and D.~Wyler, \emph{Effective lagrangian analysis of new
  interactions and flavour conservation},
  \href{http://dx.doi.org/https://doi.org/10.1016/0550-3213(86)90262-2}{\emph{Nuclear
  Physics B} {\bf 268} (1986) 621--653}.

\bibitem{Giudice:2007fh}
G.~F. Giudice, C.~Grojean, A.~Pomarol and R.~Rattazzi, \emph{{The
  Strongly-Interacting Light Higgs}},
  \href{http://dx.doi.org/10.1088/1126-6708/2007/06/045}{\emph{JHEP} {\bf 06}
  (2007) 045}, [\href{http://arxiv.org/abs/hep-ph/0703164}{{\tt
  hep-ph/0703164}}].

\bibitem{Grzadkowski:2010es}
B.~Grzadkowski, M.~Iskrzynski, M.~Misiak and J.~Rosiek, \emph{{Dimension-Six
  Terms in the Standard Model Lagrangian}},
  \href{http://dx.doi.org/10.1007/JHEP10(2010)085}{\emph{JHEP} {\bf 10} (2010)
  085}, [\href{http://arxiv.org/abs/1008.4884}{{\tt 1008.4884}}].

\bibitem{Alloul:2013naa}
A.~Alloul, B.~Fuks and V.~Sanz, \emph{{Phenomenology of the Higgs Effective
  Lagrangian via FEYNRULES}},
  \href{http://dx.doi.org/10.1007/JHEP04(2014)110}{\emph{JHEP} {\bf 04} (2014)
  110}, [\href{http://arxiv.org/abs/1310.5150}{{\tt 1310.5150}}].

\bibitem{Falkowski:2014tna}
A.~Falkowski and F.~Riva, \emph{{Model-independent precision constraints on
  dimension-6 operators}},
  \href{http://dx.doi.org/10.1007/JHEP02(2015)039}{\emph{JHEP} {\bf 02} (2015)
  039}, [\href{http://arxiv.org/abs/1411.0669}{{\tt 1411.0669}}].

\bibitem{Brivio:2017vri}
I.~Brivio and M.~Trott, \emph{{The Standard Model as an Effective Field
  Theory}}, \href{http://dx.doi.org/10.1016/j.physrep.2018.11.002}{\emph{Phys.
  Rept.} {\bf 793} (2019) 1--98}, [\href{http://arxiv.org/abs/1706.08945}{{\tt
  1706.08945}}].

\bibitem{Hagiwara:1993sw}
K.~Hagiwara and M.~L. Stong, \emph{{Probing the scalar sector in $ e^{+} e^{-}
  \to f \bar{f} H$ }}, \href{http://dx.doi.org/10.1007/BF01559529}{\emph{Z.
  Phys. C} {\bf 62} (1994) 99--108},
  [\href{http://arxiv.org/abs/hep-ph/9309248}{{\tt hep-ph/9309248}}].

\bibitem{Hagiwara:2000tk}
K.~Hagiwara, S.~Ishihara, J.~Kamoshita and B.~A. Kniehl, \emph{{Prospects of
  measuring general Higgs couplings at $e^{+} e^{-}$ linear colliders}},
  \href{http://dx.doi.org/10.1007/s100520000366}{\emph{Eur. Phys. J. C} {\bf
  14} (2000) 457--468}, [\href{http://arxiv.org/abs/hep-ph/0002043}{{\tt
  hep-ph/0002043}}].

\bibitem{Han:2000mi}
T.~Han and J.~Jiang, \emph{{CP violating Z Z H coupling at e+ e- linear
  colliders}}, \href{http://dx.doi.org/10.1103/PhysRevD.63.096007}{\emph{Phys.
  Rev. D} {\bf 63} (2001) 096007},
  [\href{http://arxiv.org/abs/hep-ph/0011271}{{\tt hep-ph/0011271}}].

\bibitem{Han:2005pu}
T.~Han, Y.-P. Kuang and B.~Zhang, \emph{{Anomalous gauge couplings of the Higgs
  boson at high energy photon colliders}},
  \href{http://dx.doi.org/10.1103/PhysRevD.73.055010}{\emph{Phys. Rev. D} {\bf
  73} (2006) 055010}, [\href{http://arxiv.org/abs/hep-ph/0512193}{{\tt
  hep-ph/0512193}}].

\bibitem{Biswal:2008tg}
S.~S. Biswal, D.~Choudhury, R.~M. Godbole and Mamta, \emph{{Role of
  polarization in probing anomalous gauge interactions of the Higgs boson}},
  \href{http://dx.doi.org/10.1103/PhysRevD.79.035012}{\emph{Phys. Rev. D} {\bf
  79} (2009) 035012}, [\href{http://arxiv.org/abs/0809.0202}{{\tt 0809.0202}}].

\bibitem{Dutta:2008bh}
S.~Dutta, K.~Hagiwara and Y.~Matsumoto, \emph{{Measuring the Higgs-Vector boson
  Couplings at Linear $e^{+} e^{-}$ Collider}},
  \href{http://dx.doi.org/10.1103/PhysRevD.78.115016}{\emph{Phys. Rev. D} {\bf
  78} (2008) 115016}, [\href{http://arxiv.org/abs/0808.0477}{{\tt 0808.0477}}].

\bibitem{Christensen:2010pf}
N.~D. Christensen, T.~Han and Y.~Li, \emph{{Testing CP Violation in ZZH
  Interactions at the LHC}},
  \href{http://dx.doi.org/10.1016/j.physletb.2010.08.008}{\emph{Phys. Lett. B}
  {\bf 693} (2010) 28--35}, [\href{http://arxiv.org/abs/1005.5393}{{\tt
  1005.5393}}].

\bibitem{Desai:2011yj}
N.~Desai, D.~K. Ghosh and B.~Mukhopadhyaya, \emph{{CP-violating HWW couplings
  at the Large Hadron Collider}},
  \href{http://dx.doi.org/10.1103/PhysRevD.83.113004}{\emph{Phys. Rev. D} {\bf
  83} (2011) 113004}, [\href{http://arxiv.org/abs/1104.3327}{{\tt 1104.3327}}].

\bibitem{Biswal:2012mp}
S.~S. Biswal, R.~M. Godbole, B.~Mellado and S.~Raychaudhuri, \emph{{Azimuthal
  Angle Probe of Anomalous $HWW$ Couplings at a High Energy $ep$ Collider}},
  \href{http://dx.doi.org/10.1103/PhysRevLett.109.261801}{\emph{Phys. Rev.
  Lett.} {\bf 109} (2012) 261801}, [\href{http://arxiv.org/abs/1203.6285}{{\tt
  1203.6285}}].

\bibitem{Cakir:2013bxa}
I.~T. Cakir, O.~Cakir, A.~Senol and A.~T. Tasci, \emph{{Probing Anomalous HZZ
  Couplings at the LHeC}},
  \href{http://dx.doi.org/10.1142/S0217732313501423}{\emph{Mod. Phys. Lett. A}
  {\bf 28} (2013) 1350142}, [\href{http://arxiv.org/abs/1304.3616}{{\tt
  1304.3616}}].

\bibitem{Maltoni:2013sma}
F.~Maltoni, K.~Mawatari and M.~Zaro, \emph{{Higgs characterisation via
  vector-boson fusion and associated production: NLO and parton-shower
  effects}}, \href{http://dx.doi.org/10.1140/epjc/s10052-013-2710-5}{\emph{Eur.
  Phys. J. C} {\bf 74} (2014) 2710},
  [\href{http://arxiv.org/abs/1311.1829}{{\tt 1311.1829}}].

\bibitem{Anderson:2013afp}
I.~Anderson et~al., \emph{{Constraining Anomalous HVV Interactions at Proton
  and Lepton Colliders}},
  \href{http://dx.doi.org/10.1103/PhysRevD.89.035007}{\emph{Phys. Rev. D} {\bf
  89} (2014) 035007}, [\href{http://arxiv.org/abs/1309.4819}{{\tt 1309.4819}}].

\bibitem{Kumar:2015kca}
M.~Kumar, X.~Ruan, R.~Islam, A.~S. Cornell, M.~Klein, U.~Klein et~al.,
  \emph{{Probing anomalous couplings using di-Higgs production in
  electron\textendash{}proton collisions}},
  \href{http://dx.doi.org/10.1016/j.physletb.2016.11.039}{\emph{Phys. Lett. B}
  {\bf 764} (2017) 247--253}, [\href{http://arxiv.org/abs/1509.04016}{{\tt
  1509.04016}}].

\bibitem{Boselli:2017pef}
S.~Boselli, C.~M. Carloni~Calame, G.~Montagna, O.~Nicrosini, F.~Piccinini and
  A.~Shivaji, \emph{{Higgs decay into four charged leptons in the presence of
  dimension-six operators}},
  \href{http://dx.doi.org/10.1007/JHEP01(2018)096}{\emph{JHEP} {\bf 01} (2018)
  096}, [\href{http://arxiv.org/abs/1703.06667}{{\tt 1703.06667}}].

\bibitem{Nakamura:2017ihk}
J.~Nakamura, \emph{{Polarisations of the $Z$ and $W$ bosons in the processes
  $pp \to ZH$ and $pp \to W^{\pm}_{}H$}},
  \href{http://dx.doi.org/10.1007/JHEP08(2017)008}{\emph{JHEP} {\bf 08} (2017)
  008}, [\href{http://arxiv.org/abs/1706.01816}{{\tt 1706.01816}}].

\bibitem{Li:2019evl}
H.-D. Li, C.-D. L\"u and L.-Y. Shan, \emph{{Sensitivity study of anomalous
  $HZZ$ couplings at a future Higgs factory}},
  \href{http://dx.doi.org/10.1088/1674-1137/43/10/103001}{\emph{Chin. Phys. C}
  {\bf 43} (2019) 103001}, [\href{http://arxiv.org/abs/1901.10218}{{\tt
  1901.10218}}].

\bibitem{Sahin:2019wew}
B.~\c{S}ahin, \emph{{Search for the anomalous ZZH couplings at the CLIC}},
  \href{http://dx.doi.org/10.1142/S0217732319502997}{\emph{Mod. Phys. Lett. A}
  {\bf 34} (2019) 1950299}.

\bibitem{Banerjee:2019pks}
S.~Banerjee, R.~S. Gupta, J.~Y. Reiness and M.~Spannowsky, \emph{{Resolving the
  tensor structure of the Higgs coupling to $Z$-bosons via Higgs-strahlung}},
  \href{http://dx.doi.org/10.1103/PhysRevD.100.115004}{\emph{Phys. Rev. D} {\bf
  100} (2019) 115004}, [\href{http://arxiv.org/abs/1905.02728}{{\tt
  1905.02728}}].

\bibitem{Han:2020pif}
T.~Han, D.~Liu, I.~Low and X.~Wang, \emph{{Electroweak couplings of the Higgs
  boson at a multi-TeV muon collider}},
  \href{http://dx.doi.org/10.1103/PhysRevD.103.013002}{\emph{Phys. Rev. D} {\bf
  103} (2021) 013002}, [\href{http://arxiv.org/abs/2008.12204}{{\tt
  2008.12204}}].

\bibitem{Bizon:2021rww}
W.~Bizo\'n, F.~Caola, K.~Melnikov and R.~R\"ontsch, \emph{{Anomalous couplings
  in associated VH production with Higgs boson decay to massive b quarks at
  NNLO in QCD}},
  \href{http://dx.doi.org/10.1103/PhysRevD.105.014023}{\emph{Phys. Rev. D} {\bf
  105} (2022) 014023}, [\href{http://arxiv.org/abs/2106.06328}{{\tt
  2106.06328}}].

\bibitem{Rao:2019hsp}
K.~Rao, S.~D. Rindani and P.~Sarmah, \emph{{Probing anomalous gauge-Higgs
  couplings using $Z$ boson polarization at $e^+ e^-$ colliders}},
  \href{http://dx.doi.org/10.1016/j.nuclphysb.2019.114840}{\emph{Nucl. Phys. B}
  {\bf 950} (2020) 114840}, [\href{http://arxiv.org/abs/1904.06663}{{\tt
  1904.06663}}].

\bibitem{Rao:2022olq}
K.~Rao, S.~D. Rindani, P.~Sarmah and B.~Singh, \emph{{Polarized $Z$ cross
  sections in Higgsstrahlung for the determination of anomalous $ZZH$
  couplings}},  \href{http://arxiv.org/abs/2202.10215}{{\tt 2202.10215}}.

\bibitem{Asteriadis:2022ebf}
K.~Asteriadis, F.~Caola, K.~Melnikov and R.~R\"ontsch, \emph{{Anomalous Higgs
  boson couplings in weak boson fusion production at NNLO in QCD}},
  \href{http://arxiv.org/abs/2206.14630}{{\tt 2206.14630}}.

\bibitem{Senol:2012fc}
A.~Senol, \emph{{Anomalous Higgs Couplings at the LHeC}},
  \href{http://dx.doi.org/10.1016/j.nuclphysb.2013.04.016}{\emph{Nucl. Phys. B}
  {\bf 873} (2013) 293--299}, [\href{http://arxiv.org/abs/1212.6869}{{\tt
  1212.6869}}].

\bibitem{Banerjee:2013apa}
S.~Banerjee, S.~Mukhopadhyay and B.~Mukhopadhyaya, \emph{{Higher dimensional
  operators and the LHC Higgs data: The role of modified kinematics}},
  \href{http://dx.doi.org/10.1103/PhysRevD.89.053010}{\emph{Phys. Rev. D} {\bf
  89} (2014) 053010}, [\href{http://arxiv.org/abs/1308.4860}{{\tt 1308.4860}}].

\bibitem{Craig:2014una}
N.~Craig, M.~Farina, M.~McCullough and M.~Perelstein, \emph{{Precision
  Higgsstrahlung as a Probe of New Physics}},
  \href{http://dx.doi.org/10.1007/JHEP03(2015)146}{\emph{JHEP} {\bf 03} (2015)
  146}, [\href{http://arxiv.org/abs/1411.0676}{{\tt 1411.0676}}].

\bibitem{Amar:2014fpa}
G.~Amar, S.~Banerjee, S.~von Buddenbrock, A.~S. Cornell, T.~Mandal, B.~Mellado
  et~al., \emph{{Exploration of the tensor structure of the Higgs boson
  coupling to weak bosons in $e^{+} e^{−}$ collisions}},
  \href{http://dx.doi.org/10.1007/JHEP02(2015)128}{\emph{JHEP} {\bf 02} (2015)
  128}, [\href{http://arxiv.org/abs/1405.3957}{{\tt 1405.3957}}].

\bibitem{Englert:2014cva}
C.~Englert and M.~Spannowsky, \emph{{Effective Theories and Measurements at
  Colliders}},
  \href{http://dx.doi.org/10.1016/j.physletb.2014.11.035}{\emph{Phys. Lett. B}
  {\bf 740} (2015) 8--15}, [\href{http://arxiv.org/abs/1408.5147}{{\tt
  1408.5147}}].

\bibitem{Ellis:2014dva}
J.~Ellis, V.~Sanz and T.~You, \emph{{Complete Higgs Sector Constraints on
  Dimension-6 Operators}},
  \href{http://dx.doi.org/10.1007/JHEP07(2014)036}{\emph{JHEP} {\bf 07} (2014)
  036}, [\href{http://arxiv.org/abs/1404.3667}{{\tt 1404.3667}}].

\bibitem{Mellado:2015ehl}
B.~Mellado, L.~March and X.~Ruan, \emph{{Probing new physics in the Higgs
  sector with effective field theories at the Large Hadron Collider}},  in
  \emph{{60th Annual Conference of the South African Institute of Physics}},
  pp.~210--215, 2015.

\bibitem{Banerjee:2015bla}
S.~Banerjee, T.~Mandal, B.~Mellado and B.~Mukhopadhyaya, \emph{{Cornering
  dimension-6 $HVV$ interactions at high luminosity LHC: the role of event
  ratios}}, \href{http://dx.doi.org/10.1007/JHEP09(2015)057}{\emph{JHEP} {\bf
  09} (2015) 057}, [\href{http://arxiv.org/abs/1505.00226}{{\tt 1505.00226}}].

\bibitem{Dwivedi:2015nta}
S.~Dwivedi, D.~K. Ghosh, B.~Mukhopadhyaya and A.~Shivaji, \emph{{Constraints on
  CP-violating gauge-Higgs operators}},
  \href{http://dx.doi.org/10.1103/PhysRevD.92.095015}{\emph{Phys. Rev. D} {\bf
  92} (2015) 095015}, [\href{http://arxiv.org/abs/1505.05844}{{\tt
  1505.05844}}].

\bibitem{Englert:2015hrx}
C.~Englert, R.~Kogler, H.~Schulz and M.~Spannowsky, \emph{{Higgs coupling
  measurements at the LHC}},
  \href{http://dx.doi.org/10.1140/epjc/s10052-016-4227-1}{\emph{Eur. Phys. J.
  C} {\bf 76} (2016) 393}, [\href{http://arxiv.org/abs/1511.05170}{{\tt
  1511.05170}}].

\bibitem{Craig:2015wwr}
N.~Craig, J.~Gu, Z.~Liu and K.~Wang, \emph{{Beyond Higgs Couplings: Probing the
  Higgs with Angular Observables at Future e$^{+}$ e$^{−}$ Colliders}},
  \href{http://dx.doi.org/10.1007/JHEP03(2016)050}{\emph{JHEP} {\bf 03} (2016)
  050}, [\href{http://arxiv.org/abs/1512.06877}{{\tt 1512.06877}}].

\bibitem{Dwivedi:2016xwm}
S.~Dwivedi, D.~K. Ghosh, B.~Mukhopadhyaya and A.~Shivaji, \emph{{Distinguishing
  $CP$-odd couplings of the Higgs boson to weak boson pairs}},
  \href{http://dx.doi.org/10.1103/PhysRevD.93.115039}{\emph{Phys. Rev. D} {\bf
  93} (2016) 115039}, [\href{http://arxiv.org/abs/1603.06195}{{\tt
  1603.06195}}].

\bibitem{Ferreira:2016jea}
F.~Ferreira, B.~Fuks, V.~Sanz and D.~Sengupta, \emph{{Probing ${CP}$-violating
  Higgs and gauge-boson couplings in the Standard Model effective field
  theory}}, \href{http://dx.doi.org/10.1140/epjc/s10052-017-5226-6}{\emph{Eur.
  Phys. J. C} {\bf 77} (2017) 675},
  [\href{http://arxiv.org/abs/1612.01808}{{\tt 1612.01808}}].

\bibitem{Degrande:2016dqg}
C.~Degrande, B.~Fuks, K.~Mawatari, K.~Mimasu and V.~Sanz, \emph{{Electroweak
  Higgs boson production in the standard model effective field theory beyond
  leading order in QCD}},
  \href{http://dx.doi.org/10.1140/epjc/s10052-017-4793-x}{\emph{Eur. Phys. J.
  C} {\bf 77} (2017) 262}, [\href{http://arxiv.org/abs/1609.04833}{{\tt
  1609.04833}}].

\bibitem{Denizli:2017pyu}
H.~Denizli and A.~Senol, \emph{{Constraints on Higgs effective couplings in
  $H\nu \bar{\nu}$ production of CLIC at 380 GeV}},
  \href{http://dx.doi.org/10.1155/2018/1627051}{\emph{Adv. High Energy Phys.}
  {\bf 2018} (2018) 1627051}, [\href{http://arxiv.org/abs/1707.03890}{{\tt
  1707.03890}}].

\bibitem{Khanpour:2017cfq}
H.~Khanpour and M.~Mohammadi~Najafabadi, \emph{{Constraining Higgs boson
  effective couplings at electron-positron colliders}},
  \href{http://dx.doi.org/10.1103/PhysRevD.95.055026}{\emph{Phys. Rev. D} {\bf
  95} (2017) 055026}, [\href{http://arxiv.org/abs/1702.00951}{{\tt
  1702.00951}}].

\bibitem{Hesari:2018ssq}
H.~Hesari, H.~Khanpour and M.~Mohammadi~Najafabadi, \emph{{Study of Higgs
  Effective Couplings at Electron-Proton Colliders}},
  \href{http://dx.doi.org/10.1103/PhysRevD.97.095041}{\emph{Phys. Rev. D} {\bf
  97} (2018) 095041}, [\href{http://arxiv.org/abs/1805.04697}{{\tt
  1805.04697}}].

\bibitem{Banerjee:2018bio}
S.~Banerjee, C.~Englert, R.~S. Gupta and M.~Spannowsky, \emph{{Probing
  Electroweak Precision Physics via boosted Higgs-strahlung at the LHC}},
  \href{http://dx.doi.org/10.1103/PhysRevD.98.095012}{\emph{Phys. Rev. D} {\bf
  98} (2018) 095012}, [\href{http://arxiv.org/abs/1807.01796}{{\tt
  1807.01796}}].

\bibitem{Karadeniz:2019upm}
O.~Karadeniz, A.~Senol, K.~Y. Oyulmaz and H.~Denizli, \emph{{CP-violating
  Higgs-gauge boson couplings in $H\nu \bar{\nu}$ production at three energy
  stages of CLIC}},
  \href{http://dx.doi.org/10.1140/epjc/s10052-020-7740-1}{\emph{Eur. Phys. J.
  C} {\bf 80} (2020) 229}, [\href{http://arxiv.org/abs/1909.08032}{{\tt
  1909.08032}}].

\bibitem{Cirigliano:2019vfc}
V.~Cirigliano, A.~Crivellin, W.~Dekens, J.~de~Vries, M.~Hoferichter and
  E.~Mereghetti, \emph{{CP Violation in Higgs-Gauge Interactions: From Tabletop
  Experiments to the LHC}},
  \href{http://dx.doi.org/10.1103/PhysRevLett.123.051801}{\emph{Phys. Rev.
  Lett.} {\bf 123} (2019) 051801}, [\href{http://arxiv.org/abs/1903.03625}{{\tt
  1903.03625}}].

\bibitem{Freitas:2019hbk}
F.~F. Freitas, C.~K. Khosa and V.~Sanz, \emph{{Exploring the standard model EFT
  in VH production with machine learning}},
  \href{http://dx.doi.org/10.1103/PhysRevD.100.035040}{\emph{Phys. Rev. D} {\bf
  100} (2019) 035040}, [\href{http://arxiv.org/abs/1902.05803}{{\tt
  1902.05803}}].

\bibitem{Henning:2019vjr}
B.~Henning, D.~M. Lombardo and F.~Riva, \emph{{Improved BSM sensitivity in
  diboson processes at linear colliders}},
  \href{http://dx.doi.org/10.1140/epjc/s10052-020-7695-2}{\emph{Eur. Phys. J.
  C} {\bf 80} (2020) 220}, [\href{http://arxiv.org/abs/1909.01937}{{\tt
  1909.01937}}].

\bibitem{Biswas:2021qaf}
T.~Biswas, A.~Datta and B.~Mukhopadhyaya, \emph{{Following the trail of new
  physics via the vector boson fusion Higgs boson signal at the Large Hadron
  Collider}}, \href{http://dx.doi.org/10.1103/PhysRevD.105.055028}{\emph{Phys.
  Rev. D} {\bf 105} (2022) 055028},
  [\href{http://arxiv.org/abs/2107.05503}{{\tt 2107.05503}}].

\bibitem{LHeCStudyGroup:2012zhm}
{\scshape LHeC Study Group} collaboration, J.~L. Abelleira~Fernandez et~al.,
  \emph{{A Large Hadron Electron Collider at CERN: Report on the Physics and
  Design Concepts for Machine and Detector}},
  \href{http://dx.doi.org/10.1088/0954-3899/39/7/075001}{\emph{J. Phys. G} {\bf
  39} (2012) 075001}, [\href{http://arxiv.org/abs/1206.2913}{{\tt 1206.2913}}].

\bibitem{Bruening:2013bga}
O.~Bruening and M.~Klein, \emph{{The Large Hadron Electron Collider}},
  \href{http://dx.doi.org/10.1142/S0217732313300115}{\emph{Mod. Phys. Lett. A}
  {\bf 28} (2013) 1330011}, [\href{http://arxiv.org/abs/1305.2090}{{\tt
  1305.2090}}].

\bibitem{Han:2009pe}
T.~Han and B.~Mellado, \emph{{Higgs Boson Searches and the $H b \bar{b}$
  Coupling at the LHeC}},
  \href{http://dx.doi.org/10.1103/PhysRevD.82.016009}{\emph{Phys. Rev. D} {\bf
  82} (2010) 016009}, [\href{http://arxiv.org/abs/0909.2460}{{\tt 0909.2460}}].

\bibitem{LHeC:2020van}
{\scshape LHeC, FCC-he Study Group} collaboration, P.~Agostini et~al.,
  \emph{{The Large Hadron-Electron Collider at the HL-LHC}},
  \href{http://dx.doi.org/10.1088/1361-6471/abf3ba}{\emph{J. Phys. G} {\bf 48}
  (2021) 110501}, [\href{http://arxiv.org/abs/2007.14491}{{\tt 2007.14491}}].

\bibitem{Benedikt:2022kan}
M.~Benedikt et~al., \emph{{Future Circular Hadron Collider FCC-hh: Overview and
  Status}},  \href{http://arxiv.org/abs/2203.07804}{{\tt 2203.07804}}.

\bibitem{Alwall:2014hca}
J.~Alwall, R.~Frederix, S.~Frixione, V.~Hirschi, F.~Maltoni, O.~Mattelaer
  et~al., \emph{{The automated computation of tree-level and next-to-leading
  order differential cross sections, and their matching to parton shower
  simulations}}, \href{http://dx.doi.org/10.1007/JHEP07(2014)079}{\emph{JHEP}
  {\bf 07} (2014) 079}, [\href{http://arxiv.org/abs/1405.0301}{{\tt
  1405.0301}}].

\bibitem{Alloul:2013bka}
A.~Alloul, N.~D. Christensen, C.~Degrande, C.~Duhr and B.~Fuks,
  \emph{{FeynRules 2.0 - A complete toolbox for tree-level phenomenology}},
  \href{http://dx.doi.org/10.1016/j.cpc.2014.04.012}{\emph{Comput. Phys.
  Commun.} {\bf 185} (2014) 2250--2300},
  [\href{http://arxiv.org/abs/1310.1921}{{\tt 1310.1921}}].

\bibitem{CMS:2018uag}
{\scshape CMS} collaboration, A.~M. Sirunyan et~al., \emph{{Combined
  measurements of Higgs boson couplings in proton\textendash{}proton collisions
  at $\sqrt{s}=13\,\text {Te}\text {V} $}},
  \href{http://dx.doi.org/10.1140/epjc/s10052-019-6909-y}{\emph{Eur. Phys. J.
  C} {\bf 79} (2019) 421}, [\href{http://arxiv.org/abs/1809.10733}{{\tt
  1809.10733}}].

\bibitem{Robson:2018zje}
A.~Robson and P.~Roloff, \emph{{Updated CLIC luminosity staging baseline and
  Higgs coupling prospects}},  \href{http://arxiv.org/abs/1812.01644}{{\tt
  1812.01644}}.

\bibitem{ATLAS:2020evk}
{\scshape ATLAS} collaboration, G.~Aad et~al., \emph{{Test of cp invariance in
  vector-boson fusion production of the higgs boson in the $h$ $\rightarrow$
  $\tau$ $\tau$ channel in proton $-$ proton collisions at $\sqrt{s}$ = 13 tev
  with the atlas detector}},
  \href{http://dx.doi.org/10.1016/j.physletb.2020.135426}{\emph{Phys. Lett. B}
  {\bf 805} (2020) 135426}, [\href{http://arxiv.org/abs/2002.05315}{{\tt
  2002.05315}}].

\bibitem{Workman:2022}
{\scshape Particle Data Group} collaboration, R.~Workman et~al., \emph{{Review
  of Particle Physics}}, .

\end{thebibliography}\endgroup

\end{document}